%% file: main_sim_revised.tex
\documentclass[10pt, letterpaper]{article}

\usepackage{authblk}
\usepackage{mathtools}
\usepackage{parskip}
\usepackage{fullpage}
\usepackage{times}
\usepackage{latexsym}
\usepackage{nameref}
\usepackage{enumitem}
\usepackage{pdfpages}
\usepackage{standalone}
\usepackage{float}
\usepackage{graphicx}
\usepackage{amssymb}
\usepackage{booktabs}
\usepackage[font=small]{caption, subcaption}
\usepackage{multirow}
\usepackage{tabularx}
\usepackage{makecell}
\usepackage{hyperref}
\usepackage{inconsolata}
\usepackage[autostyle=true]{csquotes}
\usepackage{xparse}
\usepackage[numbers]{natbib}
\usepackage{xcolor, colortbl}
\usepackage[official]{eurosym}
\usepackage{xurl}
\usepackage[]{mdframed}

\hypersetup{
	colorlinks=true,
	citecolor=blue,
	linkcolor=blue,
	urlcolor  = blue
}

\emergencystretch=\maxdimen
\makeatletter
\renewcommand\AB@affilsepx{\hspace{1em} \protect\Affilfont}
\makeatother

\renewcommand*{\Affilfont}{\fontsize{8.5}{10.5}\selectfont\normalfont}

\newcommand*\samethanks[1][\value{footnote}]{\footnotemark[1]}

\author[1-3\thanks{Corresponding authors. Email: \href{mailto:manuel.tonneau@oii.ox.ac.uk}{\nolinkurl{manuel.tonneau@oii.ox.ac.uk}}; \href{mailto:paul.rottger@oii.ox.ac.uk}{\nolinkurl{paul.rottger@oii.ox.ac.uk}}; \href{mailto:sfraiberger@worldbank.org}{\nolinkurl{sfraiberger@worldbank.org}}.}]{Manuel~Tonneau}
\author[1]{Dylan~Thurgood}
\author[4]{Diyi~Liu}
\author[2]{Niyati~Malhotra}
\author[2]{Victor~Orozco-Olvera}
\author[1]{Ralph~Schroeder}
\author[1]{Scott~A.~Hale}
\author[5]{Manoel~Horta~Ribeiro}
\author[1\samethanks]{Paul~Röttger}
\author[2, 3\samethanks]{Samuel~P.~Fraiberger}

\affil[1]{University of Oxford}
\affil[2]{World Bank Group}
\affil[3]{New York University}
\affil[4]{University of Copenhagen}
\affil[5]{Princeton University}

\renewenvironment{abstract}
 {\small
  \begin{center}
  \bfseries \abstractname\vspace{-0.5em}
  \end{center}
  \list{}{%
    \setlength{\leftmargin}{6.25mm}%
    \setlength{\rightmargin}{\leftmargin}%
  }%
  \item\relax}
 {\endlist}

\newcommand{\beginsupplement}{
  \clearpage
  \captionsetup{labelsep=period}
  \renewcommand{\figurename}{Fig.}
  \setcounter{section}{0}
  \setcounter{subsection}{0}
  \setcounter{subsubsection}{0}
  \setcounter{figure}{0}
  \setcounter{table}{0}
  \setcounter{equation}{0}
  \renewcommand\thesection{S}
  \renewcommand\thesubsection{S\arabic{subsection}}
  \renewcommand\thesubsubsection{S\arabic{subsection}.\arabic{subsubsection}}
  \renewcommand\thefigure{S\arabic{figure}}
  \renewcommand\thetable{S\arabic{table}}
  \renewcommand\theequation{S\arabic{equation}}
}

\title{The Enforcement and Feasibility of Hate Speech Moderation}
\date{}

\begin{document}

\maketitle

\vspace{-24pt}

\begin{abstract}
\noindent Online hate speech is associated with harms ranging from deteriorating mental health to violence, yet how consistently platforms moderate hate, and whether enforcement is feasible at scale, remain poorly understood. We audit hate speech moderation on Twitter (now X) using 540,000 tweets annotated by trained native speakers, representative of a full day on the platform. Five months after posting, 80\% of hateful tweets, including violent ones, remained online. Removal was only marginally more likely than for non-hateful tweets, far below scams or adult content, and insensitive to severity and reach. Automated detection could not reliably classify hate but ranked it highly, enabling human triage. Simulating this workflow, current staffing curbed little exposure, yet substantial reductions proved financially feasible, far below applicable regulatory fines. Persistent hate reflects resource allocation, not technical limits.
\end{abstract}
\vspace{-16pt}

\section*{Introduction}

Digital platforms mediate speech for billions of people worldwide, shaping how information circulates across economic, political, and social life \citep{bond2012facebook,tufekci2017twitter,enikolopov2020social,lee2025can}. Hate speech remains a persistent feature of this environment: global surveys suggest that nearly two-thirds of internet users often encounter it online \citep{unesco_ipsos_2023_disinformation_hate_speech}. Such exposure is associated with harms beyond online spaces, including deteriorating mental health \citep{braghieri2022social} and violence ranging from attacks on refugees in Germany to the Rohingya genocide in Myanmar \citep{muller2021fanning,whitten2020myanmar}.

In this context, platforms face sustained pressure from policymakers \citep{alkiviadou2019hate, gorwa2021elections} and the public \citep{solomon2024illusory,munzert2025citizen} to take action against hate speech. Their primary instrument is content moderation \citep{roberts2017content}, the systematic screening of user-generated content by artificial intelligence (AI) classifiers and human reviewers to assess policy violations and determine appropriate actions \citep{gillespie2022not}. Despite moderation's central role for user safety and free expression \citep{gillespie2018custodians}, two fundamental questions remain unresolved: how consistently do platforms enforce their own hate speech policies, and do observed enforcement gaps reflect the technical limits of moderation at scale?

Empirical evidence on the enforcement of hate speech policies remains limited by structural barriers to auditing moderation systems. Platforms provide partial and sometimes misleading information about moderation practices \citep{giansiracusa2021facebook}, while academic research is constrained by restricted access to platform data \citep{leerssen2023end}. As a result, most studies rely on convenience samples that capture only fragments of enforcement, such as deleted tweets \citep{almuhimedi2013tweets,bhattacharya2016characterizing,zhou2016tweet} or suspended users \citep{chowdhury2020twitter,pierri2023does} drawn from narrow, non-representative user groups. The lack of representative data also complicates the evaluation of whether stronger enforcement is feasible, preventing reliable estimates of real-world detection performance and of the enforcement achievable under alternative levels of moderation staffing.

To address these gaps, we conducted what is, to our knowledge, the first global audit of both the enforcement and feasibility of hate speech moderation on a major social media platform. We focused on Twitter prior to its late 2022 acquisition, when sustained public data access enabled large-scale external auditing. We used a complete snapshot of public tweets posted over a 24-hour period on September 21, 2022, collected 10 minutes after publication \citep{pfeffer2023just}. From this dataset, we drew representative samples stratified across eight major platform languages and four English-dominant countries, capturing both linguistic and geographic variation (Fig.~\ref{fig:summary_diagram}). We then conducted large-scale manual annotation to identify content characteristics, including whether tweets contain hate speech as defined by Twitter’s Hateful Conduct policy at the time (materials and methods). The resulting dataset contains 1.6 million labels for 540,000 tweets and is the largest manual annotation effort in hate speech research to date.

Using this dataset, we analyzed moderation along two dimensions. First, we examined enforcement by querying tweets and accounts five months after posting to measure hateful content removal and user suspension rates. Second, we assessed feasibility by evaluating automated hate speech detection systems and simulating human–AI moderation pipelines to estimate achievable reductions in exposure to hate speech. Together, these analyses provide a unified assessment of how hate speech policies were enforced and whether persistent online hate reflects technical constraints or institutional choices in allocating moderation resources.

\input{figures/summary_diagram}

\section*{Results} \label{Results}

We report results at the language level throughout. Because country assignment relies on self-reported profile locations, which are only available for about a quarter of users, the country samples are less representative than the language samples; we use them as a secondary check on geographic generalizability and report those estimates in the supplementary material.

Although hate speech accounts for a small share of tweets, the platform's scale translates this rate into large absolute volumes and routine user exposure. Using the annotated sample from our 24-hour snapshot, we estimated that 0.29\% of tweets were hateful and 0.025\% were violent hate speech on average across languages (\ref{si_prevalence_composition_stats}). Extrapolating across the eight study languages, this corresponded to roughly 225,000 hateful original tweets per day (95\% confidence interval, CI [202,195, 249,684]) and 11,000 violent ones (95\% CI [7,657, 15,205]). Based on typical usage patterns, users would on average encounter roughly one hateful tweet every two days and one violent hateful tweet every three weeks in a chronological feed (\ref{si_exposure_estimation}). Exposure is likely higher in feeds optimized for engagement, which have been shown to amplify divisive content \citep{milli2025engagement}.


\input{figures/low_enforcement}

We next examined how Twitter's hate speech policies were enforced, by re-querying the annotated records five months after their initial collection. Enforcement was limited: only 20.4\% of hateful tweets had been removed (Fig.~\ref{fig:low_enforcement}A; breakdown by language in table~\ref{tab_si:count_removal_hate}).  When focusing on user suspensions, which are unambiguously the result of platform actions, the rate fell to 8.6\%, with only 8.4\% of users who posted hate speech being suspended.

Hate speech was removed only slightly more often than non-hateful tweets, and far less often than content in other policy categories. Entered jointly with the other categories, as plotted in Fig.~\ref{fig:low_enforcement}B, violent and non-violent hate each carried a removal premium indistinguishable from zero (neither $p<0.10$), whereas scam and adult content carried premiums of $+29$ and $+17$ points (both $p<0.001$; materials and methods, \ref{si_reg_spec_tables}). The hate premium was small under every specification we estimated, ranging from $+0.2$ to $+2.8$ percentage points across control sets, an order of magnitude below the scam premium; on the full language sample it was $+2.6$ points (95\% CI $[-0.4, +5.6]$), with an equivalence test rejecting premiums larger than $\pm5.2$ points (table~\ref{tab_si:removal_equivalence_power}). Account suspension showed the same pattern ($\beta=-0.005$, $p=0.55$; table~\ref{tab_si:user_suspension_language_reg}).

What little removal occurred was largely unresponsive to severity and reach. Violent hate was no more likely to be removed than hate without explicit violence ($-1.8$ percentage points with controls, $p=0.76$; \ref{si_reg_spec_tables}), with several tweets openly calling for mass violence against protected groups remaining online (\ref{si_enforcement_examples_unremoved_violent}). Nor did removal rise with a tweet's engagement, our proxy for exposure (Fig.~\ref{fig:low_enforcement}C). If anything the relationship ran the other way: the interaction between hate and log engagement was negative in the sample weighted by engagement ($-0.04$, $p<0.05$; table~\ref{tab_si:engagement_removal_interaction}), so the hateful tweets users were most likely to see were the least likely to be removed. Removal was also unrelated to how confidently annotators flagged content as hateful (\ref{si_annotator_agreement_removal}) or to the language of the content (\ref{si_reg_spec_tables}). Because hateful tweets are rare, the severity, agreement, and cross-context tests detected only sizable differences, so we read those nulls as an absence of detectable prioritization rather than proof of none.

Finally, we found no evidence that visibility reduction (``downranking'') limited the reach of the most severe hate. Twitter's policy lists downranking as a form of moderation \citep{twitter2022hateful}, which we cannot observe directly without view count data. If severe hate were consistently downranked, however, it should accrue less engagement than comparable content. Instead, the effect of violent hate on early engagement was positive and not significant (table~\ref{tab_si:engagement_language}).

Together, these results reveal substantial gaps between the stated goals of Twitter's hate speech policy and its enforcement. Whether these gaps reflect the technical limits of detection at scale or the resources platforms devote to moderation is the question we turn to next.


\input{figures/ai_perf}

 We next examined whether higher moderation rates could be achieved with AI-based hate speech detection. We evaluated a fully automated setting, in which any tweet an AI model flagged as hateful would be moderated. We assessed widely used open-source supervised models and the commercial Perspective API, selecting the best per language. We also evaluated GPT-5.1, a leading large language model released after data collection, in a zero-shot setting (materials and methods).

Across all languages, fully automated moderation would either over-moderate large amounts of non-hateful content or fail to moderate a substantial share of hate speech. Indeed, no classification threshold achieved both high precision and high recall (Fig.~\ref{fig:ai_perf}A). Increasing recall substantially reduced precision, while improving precision sharply lowered recall. GPT-5.1 exhibited a similar pattern, achieving high recall only at the cost of very low precision. Model performance was particularly poor for non-European languages, while performance for English, though generally better resourced, was worse than in several other European languages (\ref{si_ai_detection}).

To understand the sources of these errors, we examined tweets assigned the highest hate scores by the best supervised learning models (Fig.~\ref{fig:ai_perf}B), focusing on non-hateful tweets misclassified as hateful. Low precision stemmed mainly from offensive content, which used profane or aggressive language without targeting protected groups and so superficially resembled hate speech (38\% of the top 0.5\% of tweets by hate score on average, \ref{si_ai_error_analysis}). Models also misclassified tweets that quoted or mentioned hate rather than expressed it, and cases that depended on context or spelling (\ref{si_ai_error_analysis}). Precision therefore stayed low even when models flagged only their highest-scoring tweets, and lowering the score cutoff to catch more hate filled the output with false positives.

Despite these limitations, supervised models consistently ranked hateful content near the top of the score distribution (Fig.~\ref{fig:ai_perf}C). Across languages, approximately half of hateful tweets fell within the top 3.2\% of scored content. This ranking ability makes them suitable for moderation triage, focusing scarce reviewer time on the content most likely to violate policy. The binding question therefore shifts from whether hate can be detected to how much human review capacity is available to act on what is surfaced, which we examine next.


\input{figures/human_in_the_loop_moderation_revised}

We simulated a human–AI moderation pipeline of the kind large platforms use \citep{barrett2020moderates, avadhanula2022bandits}, examining how the number of human reviewers affects moderation outcomes. Because harm arises when users see hateful content, we took the reduction in exposure to hate as our primary outcome, using early engagement as a proxy. Given that removal was low overall and did not increase with a tweet's engagement (Fig.~\ref{fig:low_enforcement}), we expected this reduction to be low under current staffing. Tweets were ranked by a score combining predicted hatefulness and engagement within ten minutes of posting, with moderators reviewing the highest scored content first, as in a triage system. So that our estimates represent an upper bound on the human resources required, we assumed a more resource-intensive review process than platforms likely use (materials and methods, \ref{si_human_ai_formal}).

With reported moderator counts \citep{X_DSA_Transparency_Report_Nov2023} and the best publicly available models, moderation prevented only 24.5\% of exposure on average across the six languages with reported staffing (Fig.~\ref{fig:human-in-the-loop}A). Under the assumptions of our simulation, existing moderation capacity was therefore insufficient to substantially reduce exposure to hate. Performance also varied substantially across languages. German content reached high levels of exposure reduction with roughly 1,000 moderators, whereas Arabic required at least an order of magnitude more moderators to achieve comparable levels. These differences reflected variation in both content volume and detection performance. Because moderation resources were not allocated in proportion to these differences, exposure reductions varied widely: under reported staffing levels, English, French, and German achieved meaningful reductions in exposure to hate speech, whereas Arabic, Spanish, and Portuguese remained effectively unmoderated.

To assess feasibility, we priced moderation effort, assuming an hourly wage of USD 20 for human moderators, adjusted for purchasing power (materials and methods). We then compared the annual cost of substantially reducing exposure to hate speech with the maximum fines set by content moderation regulations (Fig.~\ref{fig:human-in-the-loop}B). Across all eight languages combined, avoiding 80\% of exposure cost 7.9\% of global revenue, below the 10\% Online Safety Act ceiling though not significantly so ($p=0.18$; \ref{si_cost_analysis}). This aggregate, however, prices far more content than any single regime governs. We therefore focused on English and German, the two languages we cover whose content is subject to a regulation that fines platforms for systematically failing to remove illegal content such as hate speech: the United Kingdom's Online Safety Act (OSA) and, at the time of our data collection, Germany's Network Enforcement Act (NetzDG). For English, the largest platform language, avoiding 80\% of exposure to hate speech cost only about 1.3\% (95\% CI [0.3, 1.7]) of global annual revenue, nearly an order of magnitude below the 10\% of global turnover that the OSA sets as its maximum penalty. Since this figure prices moderation of all English-language content worldwide, it is a conservative upper bound on the resources required for UK-specific compliance. For German, achieving the same reduction would cost approximately USD 21 million (95\% CI [8, 30] million), less than half of NetzDG's fixed EUR 50 million maximum fine, which unlike the OSA cap is an absolute amount. Substantially reducing exposure to hate speech through human--AI moderation is therefore financially feasible: for the content each regime actually governs, it costs far below the maximum penalties platforms face.

\section*{Discussion} \label{discussion}

Twitter's enforcement of hate speech policies was limited and misaligned with its stated priorities. Five months after posting, only about 20\% of hateful tweets had been removed. The only prior Twitter estimate, limited to English, found an even lower removal rate of 3.6\% one month after posting in 2020 \citep{jimenez2023economics}, indicating that our low-enforcement finding is not an artifact of the single day we sampled. Low removal is not specific to Twitter either: leaked documents indicate Facebook removed only about 5\% of hateful content in 2021 \citep{giansiracusa2021facebook}, pointing to a broader pattern of limited hate speech enforcement. 

Moreover, hateful tweets were removed only marginally more often than non-hateful ones, an order of magnitude below the premiums other policy categories carried and in sharp contrast to Twitter's last pre-acquisition transparency report, covering the second half of 2021, which presented hate speech as one of the most actioned categories \citep{twitter2022transparency}. This suggests that the removals we observed stemmed largely from reasons unrelated to hatefulness. Finally, even within hateful content, removal was largely unresponsive to severity and, if anything, inversely related to visibility, a pattern inconsistent with harm-based prioritization.

Against this backdrop, our analysis underscores both the potential and the limitations of AI-based moderation. Unlike on curated benchmarks \citep{arango2019hate,wiegand-etal-2019-detection}, no classifier we evaluated on representative data could identify hate reliably without either moderating too much benign content or missing much of it \citep{tonneau-etal-2025-hateday}. Prior work attributes this to the difficulty of separating offensive from hateful language \citep{davidson2017automated} and use from mention \citep{gligoric-etal-2024-nlp, lee2024people}. We found that these challenges are amplified in real-world settings, where merely offensive content is far more prevalent. Improvements in training data and model design may help, but performance remains bounded by the semantic overlap and ambiguity at the boundary between hate and non-hate, where even human annotators only moderately agree.

Nevertheless, these models rank hateful content well enough to enable effective human–AI triage, making substantial reductions in exposure to hate financially feasible, at a cost that, for the content each regime actually governs, falls far below the maximum fines platforms face. Our cost estimate is a conservative upper bound, and in practice moderator learning and richer contextual signals would lower the resources required \citep{ahmed2022tackling}. Because textual hate speech is comparably rare and rankable by current models on other large platforms such as Facebook and YouTube, this feasibility is unlikely to be specific to Twitter.

These results point to broader constraints shaping platform moderation. Expanding moderation capacity is costly, particularly because divisive and polarizing content often drives engagement \citep{rathje2021out,beknazar2025toxic}, creating tension between user protection and profit maximization \citep{beknazar2024model}. Platforms have increasingly resolved this tension by cutting moderation costs, scaling back human review in favor of AI automation \citep{latiff2024bytedance,moran2025endts}, often in the name of protecting free expression from excessive removal \citep{kaplan2025morespeech}. Yet our findings suggest this approach protects neither: by sharply increasing the error rate, moderation without human review leaves more hate online while removing more legitimate speech, failing both the users it should protect and the free expression it claims to defend. A more effective approach is to use AI for triage while retaining human decision-making, where residual errors stem from human judgment and can be minimized through multiple independent reviews coupled with targeted training and clear decision guidelines \citep{lee2024people,calabrese-etal-2024-explainability}. Error rates are likely to be especially low for more extreme forms of hate speech, where annotator agreement in our data is highest and public support for action is broadest \citep{solomon2024illusory,munzert2025citizen}. Finally, stronger enforcement need not imply broader censorship, as platforms could rely on measures such as visibility reduction to limit exposure while preserving free expression. 

These findings also carry direct implications for policymakers. Platforms do not adequately curb hate speech on their own, so regulation is needed to price the negative externality it imposes. Our cost estimates offer a benchmark for calibrating penalties: because moderating hate costs far less than the maximum fines platforms face, those fines are already high enough to make moderation the rational choice. However, a penalty shapes incentives only if it is enforced, and enforcement has lagged. For example, German content in our sample fell under NetzDG's removal mandate, the only such regime in force among the jurisdictions our study covers. Yet German hate removal was no higher than in languages under no comparable obligation (\ref{si_reg_spec_tables}), perhaps because NetzDG's fines targeted reporting and procedural failures instead of the removal duty itself \citep{netzdg_facebook_fine_2019, netzdg_telegram_fine_2022}. Such enforcement is only now emerging: in December 2025, the European Commission issued its first Digital Services Act fine, EUR 120 million against X for breaching transparency obligations, with a separate inquiry into its handling of illegal content still ongoing \citep{CommissionFinesEU120}. That such penalties are already consequential is clear from platforms' readiness to contest them: Meta is challenging in the High Court how financial penalties under the UK OSA are calculated \citep{tobin2026meta}. More broadly, these findings highlight the urgency of timely and independent assessments of sociotechnical risks. As researcher access to platform data continues to contract, maintaining oversight mechanisms is essential for transparency and accountability \citep{bak2025risks, turner2025transparency} and for identifying harms before they scale \citep{orben2025fixing}. 

This study has several limitations that highlight directions for future research. First, our analysis focuses on text-based hate speech, whereas visual and audiovisual content may require substantially different moderation approaches. Second, our estimates are based on a single day of platform activity. While our low-enforcement finding is unlikely to be an artifact of this snapshot, aligning with the only other Twitter estimate from a different period \citep{jimenez2023economics}, our feasibility estimates depend on the prevalence, composition, and engagement distribution of hateful content, which may vary over time. Third, our estimates assume timely moderation shortly after posting, which may require more reactive workflows in practice \citep{schneider2023effectiveness}. Finally, our 5-month enforcement window spans Twitter's October 2022 change in ownership, so in principle the low enforcement we document could reflect either the pre- or post-acquisition regime. However, the pre-acquisition team remained in place for roughly the first month, yet even the most evident violent hate was not promptly removed (\ref{si_kw_search}), and excluding accounts reinstated under the new ownership leaves all estimates unchanged (\ref{si_robustness_checks}). We therefore attribute the enforcement we observe predominantly to the pre-acquisition team.
Despite these limitations, our results provide a benchmark for evaluating moderation under real-world, multilingual conditions, and show that the persistence of hate speech is not an inherent limit of detection but rather reflects how platforms allocate moderation resources under competing incentives. Substantial reductions in exposure will therefore depend less on technical capacity than on regulators' ability to reshape those incentives.


\section*{References and Notes}

{
\footnotesize
\renewcommand{\refname}{}
\vspace{-2em}
\bibliography{main}
\bibliographystyle{unsrtnat}
}

\section*{Acknowledgments}

We have benefited from seminar and conference feedback at the MilaNLP Seminar, the Hertie School Data Science Brownbag Seminar, the Oxford Internet Institute’s Future of Social Media Research workshop, Meta’s CSS Seminar, AoIR Flashpoint Symposium, and the Social Media and Content Moderation workshop at Royal Holloway. This work was also supported in part through the NYU IT High Performance Computing resources, services, and staff expertise.

\paragraph{Funding:} The study was supported by funding from the United Kingdom’s Foreign Commonwealth and Development Office (FCDO), the World Bank’s Research Support Budget, and the Gates Foundation. The findings, interpretations, and conclusions expressed in this article are entirely those of the authors. They do not necessarily represent the views of the International Bank for Reconstruction and Development/World Bank and its affiliated organizations or those of the Executive Directors of the World Bank or the governments they represent.

\paragraph{Author contributions:} M.T. conceived and led the study. N.M., V.O., and S.P.F. acquired funding. M.T., D.T., D.L., R.S., S.H., M.H.R., P.R., and S.P.F. contributed to study design. M.T. collected the data, with support from D.L. and N.M. M.T. performed the statistical analyses, with support from D.T. M.T. wrote the first draft of the manuscript. All authors reviewed and edited the manuscript.

\paragraph{AI use:} Claude (Anthropic) was used as an aid in manuscript preparation, specifically formatting, copy-editing, and auxiliary scripts for file conversion and figure-data export. It was not used to conduct the analyses or to generate research findings. The authors are accountable for all content.

\paragraph{Competing interests:} The authors declare no competing interests.

\paragraph{Data, code, and materials availability:} All data and code needed to reproduce the analyses are deposited at Zenodo \citep{tonneau2026replication}. The deposit contains the 540,000 sampled tweet IDs with their human annotations, model scores, engagement metrics and five-month enforcement outcomes. In line with the Twitter/X Terms of Service, we do not redistribute raw tweet text; records are keyed by tweet ID so that text can be rehydrated by anyone with platform access, and account identifiers are replaced by salted hashes. The underlying 24-hour collection is archived at GESIS \citep{pfeffer2023twitterday_data}, and the human annotations alone are available as the HateDay dataset at \url{https://huggingface.co/datasets/manueltonneau/hateday}. 

\section*{List of Supplementary Materials}

\noindent Materials and Methods\\
Supplementary Text\\
Figs. S1 to S3\\
Tables S1 to S36\\
References (\textit{55}--\textit{65})\\
Data S1

\newpage

\input{si_final_revised}




\end{document}

%% file: figures/summary_diagram.tex

\begin{figure}[H]
  \centering
    \includegraphics[width=0.78\textwidth]{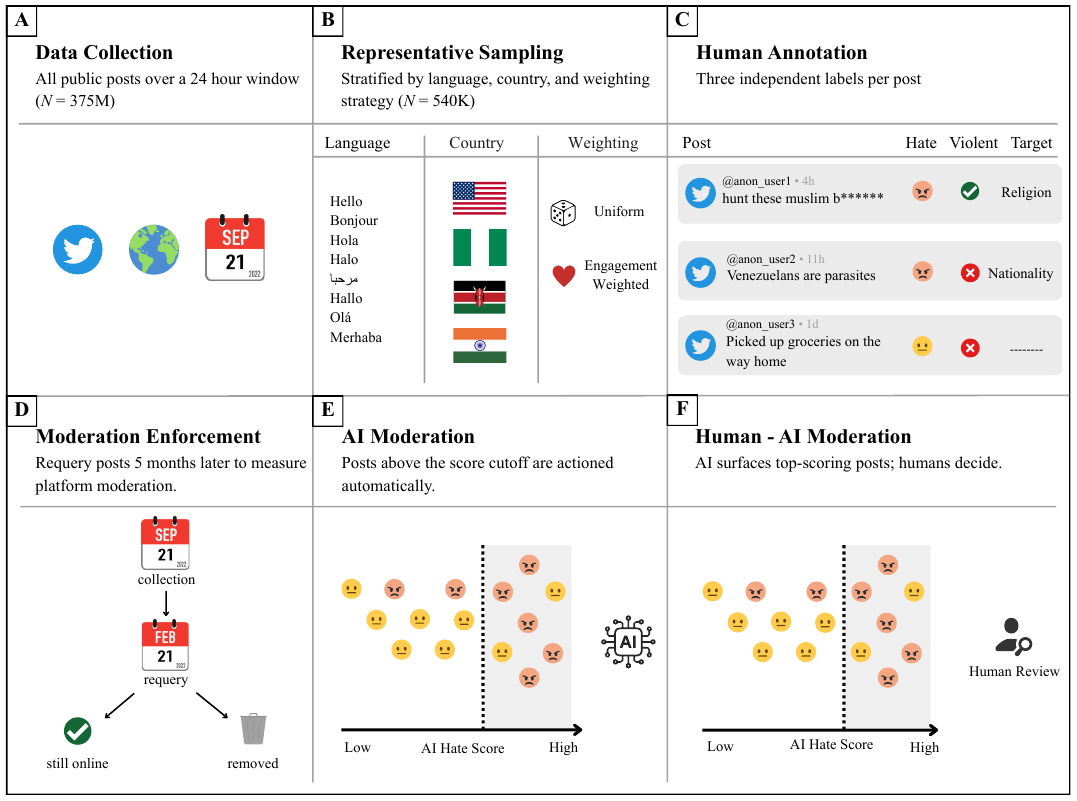}
  \caption{\textbf{Study design for auditing hate speech moderation enforcement and feasibility at platform scale.} (A) We collected all 375 million public tweets posted within a 24-hour window on September 21, 2022. (B) From this snapshot, we drew representative samples (N $=$ 540K) stratified by language, country, and weighting strategy. (C) Each sampled tweet received three independent human annotations identifying whether it is hateful, whether the hate is violent, and the targeted group. (D) To measure platform enforcement, we requeried each tweet and account five months later and recorded whether it remained online or had been removed. (E) To evaluate fully automated moderation, we assessed the performance of state-of-the-art hate speech classifiers. (F) We then simulated a human--AI pipeline in which AI classifiers score tweets and human reviewers act on the highest-scoring content. We used the simulation to estimate the workforce and cost required to reduce user exposure to hate speech.}
  \label{fig:summary_diagram}
\end{figure}

%% file: figures/low_enforcement.tex


\begin{figure}[H]
  \centering
    \includegraphics[width=\textwidth]{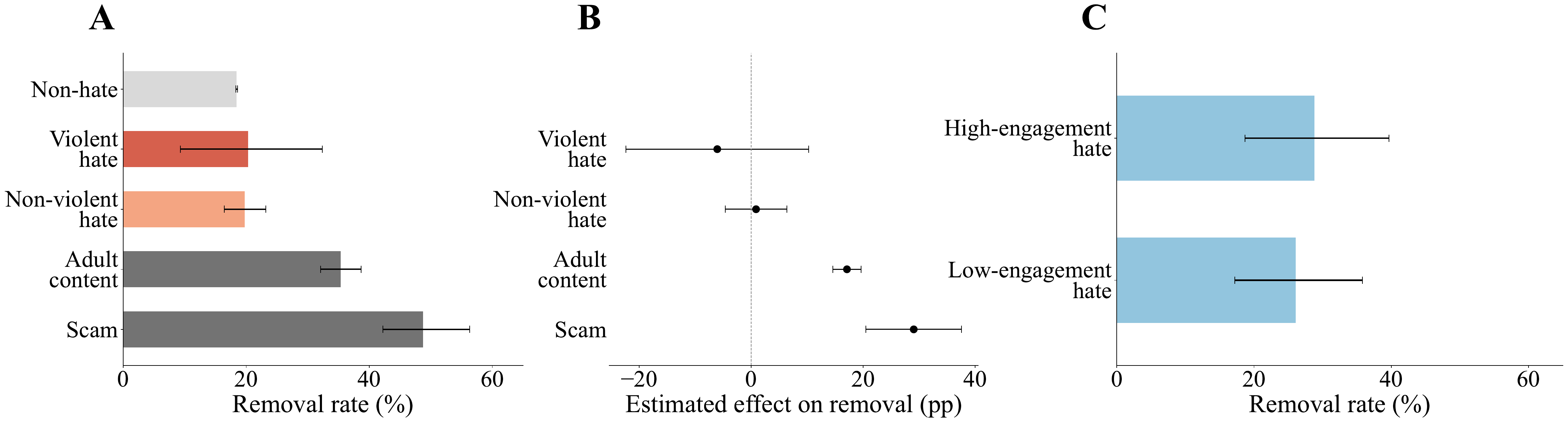}
    \caption{\textbf{Hate speech enforcement is low and unresponsive to severity and reach.} (A) Average tweet removal rates by content category across languages, with 95\% bootstrapped confidence intervals. (B) Associations between tweet characteristics and the likelihood of removal, from a linear probability model fit with indicators for hate, scam, and adult content entered jointly alongside tweet- and author-level controls and language fixed effects. Coefficients are percentage-point differences in removal probability relative to neutral tweets; horizontal lines show 95\% confidence intervals (materials and methods). (C) Removal rates of hateful tweets by early engagement, using samples weighted by engagement within each language. Tweets were split into high- and low-engagement groups at the median engagement, measured ten minutes after posting (materials and methods). Error bars show 95\% bootstrapped confidence intervals.}
  \label{fig:low_enforcement}
\end{figure}

%% file: figures/ai_perf.tex
\begin{figure}[H]
  \centering
    \includegraphics[width=\textwidth]{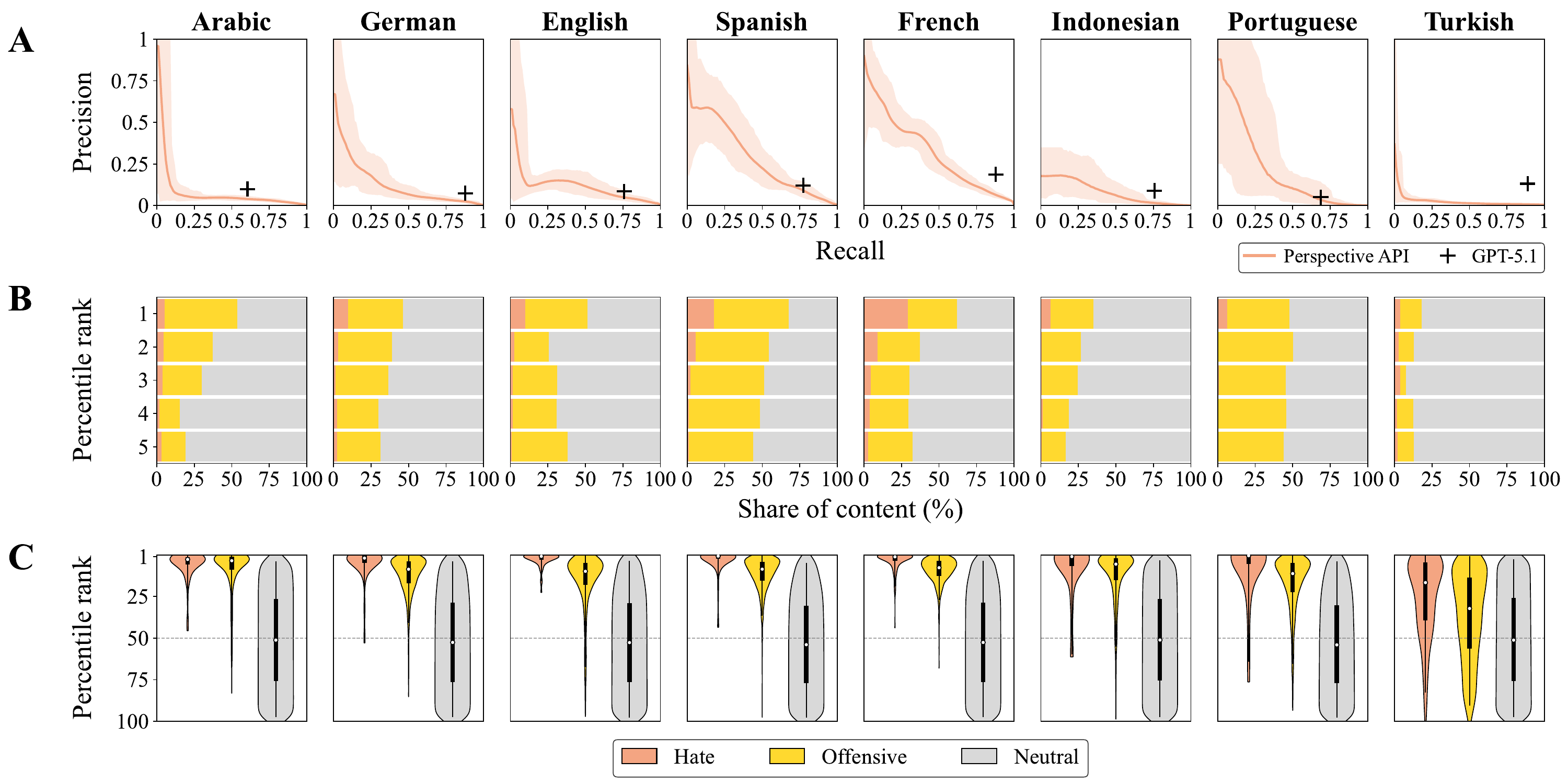}
  \caption{\textbf{AI performs poorly as a classifier but ranks hate near the top of its score distribution}. (A) Precision–recall curves of the best supervised model (Perspective API, IDENTITY\_ATTACK score) for each language, with shaded bands showing 95\% bootstrap confidence intervals. Crosses denote GPT-5.1 (zero-shot), which outputs a single discrete prediction per tweet rather than a score. (B) Share of hateful, offensive and neutral content among the top 5\% scored tweets. Each bar corresponds to a 1-percentile-point bin within the top 5\% (1 = top 1\%, 5 = 4–5\%). (C) Box plots of the percentile ranks of hateful, offensive, and neutral tweets across languages under each language's best supervised model. Lower values indicate higher scores. The dashed line marks the 50th percentile.}
  
  \label{fig:ai_perf}
\end{figure}

%% file: figures/human_in_the_loop_moderation_revised.tex
\begin{figure}[H]
  \centering
      \includegraphics[width=\textwidth]{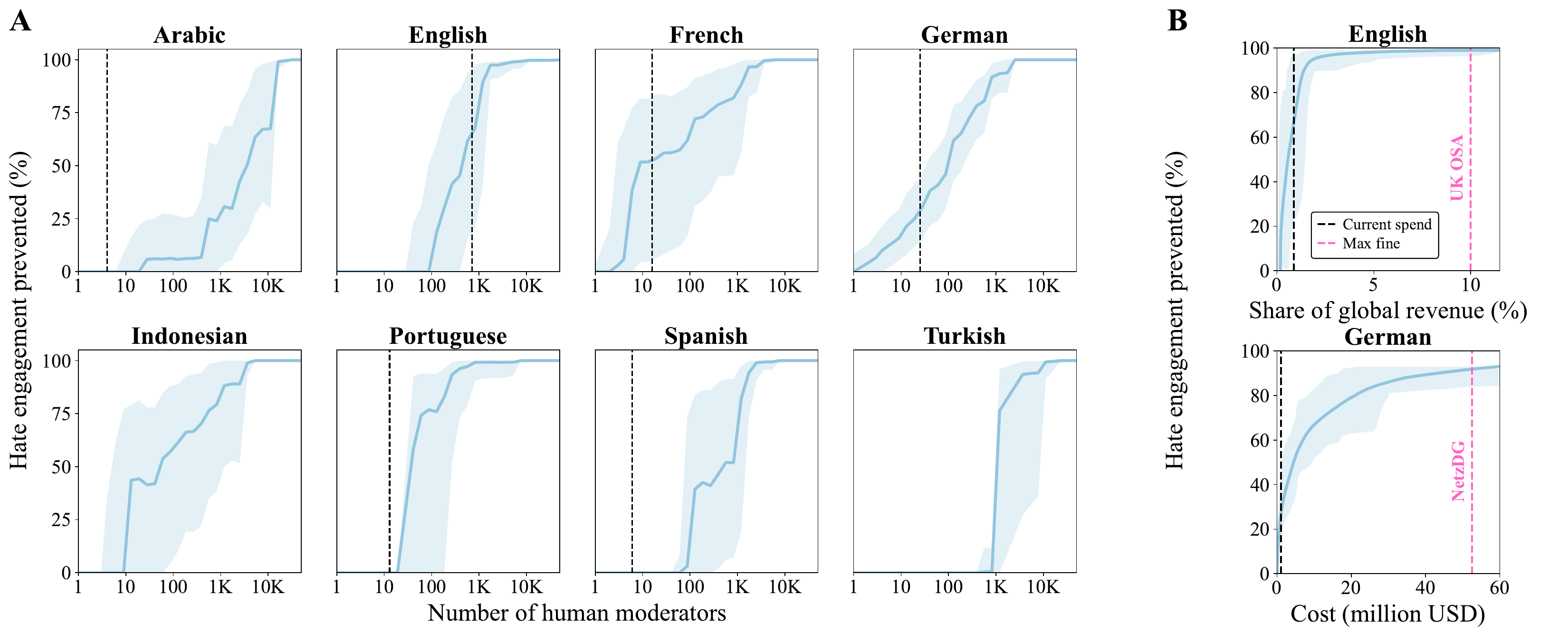}
  \caption{\textbf{Under conservative assumptions on human review, reported staffing is insufficient to achieve high moderation rates, but AI-guided review could prevent most engagement with hate at a cost below maximum regulatory fines.} (A) Share of engagement with hateful tweets, our proxy for exposure, that moderation shortly after posting would prevent, shown as a function of the number of hate speech moderators (log scale). Vertical black dashed lines mark reported moderator counts (materials and methods; Indonesian and Turkish omitted, not reported). (B) Hate engagement prevented as a function of moderation cost, for English (UK Online Safety Act, top) and German (Germany's Network Enforcement Act, NetzDG; bottom). Cost is a share of annual revenue for English, against the Act's maximum fine of 10\% of global turnover, and absolute USD for German, against NetzDG's maximum fine of EUR 50~million (both marked by pink dashed lines). Black dashed lines mark current spend implied by reported staffing. Shaded bands are 95\% bootstrap confidence intervals.}
  \label{fig:human-in-the-loop}
\end{figure}

%% file: si_final_revised.tex
\beginsupplement
\section{Supplementary Materials}

\subsection{Materials and Methods}\label{sec:mm}

\subsubsection{Ethics} All methods were approved by the University of Oxford Research Ethics Committee (reference: OII\_C1A\_23\_013). The study analyzed publicly accessible tweets and account level metadata. No private user data was collected. Annotators were compensated at hourly rates calibrated to local labor markets (\ref{si_annotation}) and were screened, briefed on potential exposure to harmful content, and provided with the option to withdraw at any time.

\subsubsection{Data collection and sampling} We analyzed a global snapshot of all tweets (N=375 million) posted on Twitter (now X) during a 24-hour period beginning on September 21, 2022, collected approximately ten minutes after they were posted as described in \citet{pfeffer2023just}. At the time of collection, the platform’s Hateful Conduct Policy stated that hateful conduct could result in tweet removal, visibility filtering, or account suspension \citep{twitter2022hateful}. From this global dataset, we selected eight languages (Arabic, English, French, German, Indonesian, Portuguese, Spanish, and Turkish) to capture broad geographic and linguistic coverage. For each language, we drew two samples: a random sample of 30{,}000 tweets and an engagement-weighted sample of 15{,}000 tweets, where the probability of selection was proportional to total early engagement (likes, replies, retweets, quote tweets) within the ten-minute collection window. To assess geographic variation independent of language, we additionally selected four countries where the main language at the time on Twitter was English (the United States, India, Nigeria, and Kenya), and drew analogous random and engagement-weighted samples for each country. Across all languages and countries, the final dataset comprises 540{,}000 tweets (\ref{si_data_sampling}).

\subsubsection{Annotation protocol}
All tweets were annotated by trained native speakers using a prescriptive protocol \citep{rottger-etal-2022-two} aligned with Twitter’s Hateful Conduct Policy in effect at the time of data collection \citep{twitter2022hateful}. Each tweet was assigned to one of three mutually exclusive categories: \textit{hate}, \textit{offensive}, or \textit{neutral}. Hate speech was defined as content that promotes or incites violence against, or directly attacks, a protected group, including groups defined by race, ethnicity, nationality, religion, gender, or sexual orientation. Offensive content was defined as abusive, profane, or insulting language that does not target a protected group and is therefore not prohibited under the policy. Neutral content included all remaining tweets that were neither hateful nor offensive.
Each tweet was independently labeled by three annotators, with final labels determined by majority vote. Inter-annotator agreement for the three-way classification was moderate (Fleiss’ $\kappa = 0.50$). All tweets marked as hateful by at least one annotator were subsequently reviewed by the first author to ensure consistent alignment with the policy definition. Tweets with a majority hate label were then further classified as \textit{violent} or \textit{non-violent} hate, where violent hate includes explicit calls for physical harm, extermination, or direct threats. Two independent annotators conducted this secondary classification and achieved substantial agreement (Cohen’s $\kappa = 0.65$ pooled across samples).
On about half of the full sample, annotators additionally labeled whether tweets contained scam or adult content, allowing comparison of hate speech enforcement with other policy-relevant content categories. Annotation guidelines, agreement statistics, and additional details are provided in \ref{si_annotation}.

\subsubsection{Enforcement measurement} 
Five months after the initial data collection, in February 2023, we assessed whether each tweet and its posting account remained accessible by querying the Twitter Compliance API. A tweet was classified as ``removed'' if it was no longer retrievable because (i) the posting account had been suspended, or (ii) the tweet itself had been deleted. Although the API provides user-level removal reasons (e.g., suspension), it does not distinguish between platform-initiated and user-initiated tweet deletion. Using our annotated representative samples, we then computed removal rates across languages and countries and estimated associations between tweet characteristics and removal or suspension likelihood using an ordinary least squares (OLS) linear probability model, whose coefficients can be read directly as percentage-point differences in removal probability. The benchmark specification underlying the comparison of hate with scam and adult content enters indicators for all three categories jointly, with tweet- and author-level controls (log retweet count, the platform's sensitive-media flag, reply status, and verified-author status) and language fixed effects (country fixed effects in the country sample), taking neutral content as the reference category. Because scam and adult content were annotated on about half of the sample, this joint model is estimated on that policy-annotated subsample. Full specifications and robustness checks are reported in \ref{si_enforcement_measurement}. Moderation actions occurring within the roughly ten-minute window between posting and collection cannot be directly observed. However, keyword-based searches for severe hate speech within the collection window revealed explicit violations that remained publicly accessible at the time of capture, suggesting that large-scale immediate removal was unlikely. Additionally, removals were assessed five months after posting, spanning a platform ownership transition. While account reinstatements under the subsequent administration could mechanically reduce observed removal rates, most enforcement likely occurred within the first weeks after posting, when the pre-acquisition moderation team remained in place. Additional details on measurement, descriptive statistics, regression specifications, and robustness analyses are provided in \ref{si_enforcement_measurement}.

\subsubsection{AI detection performance} We evaluated publicly available hate speech detection systems across languages, including open-source supervised models, the Perspective API, and the decoder-based large language model GPT-5.1 evaluated in a zero-shot setting. For each language, we selected the five most widely used open-source supervised hate speech classifiers available on Hugging Face, as determined by cumulative download counts at the time of evaluation. Perspective API was included as a widely deployed commercial toxicity detection system, and GPT-5.1 as a state-of-the-art decoder-based model released after data collection. Its inclusion assesses contemporary AI capability rather than the systems available at the time of posting.
All models were evaluated fully out-of-sample on our annotated representative datasets without additional fine-tuning. For supervised models (including Perspective API), performance was assessed using average precision (AP), which is robust to class imbalance and reflects ranking quality. For GPT-5.1, which produces discrete class predictions rather than continuous scores, we reported precision and recall. For each language, we selected the supervised model with the highest AP for subsequent analyses. To better understand systematic misclassifications, we conducted a qualitative analysis of high-confidence false positives among the top 0.5\% of model scores. Full model details and evaluation procedures are provided in \ref{si_ai_detection}. 

\subsubsection{Simulation of human--AI collaborative moderation}
We simulated a two-stage human–AI moderation pipeline, focusing on the language level. In the first stage, all tweets were assigned a continuous hatefulness score by the best-performing supervised learning model for the corresponding language, as identified by highest average precision on our representative evaluation set. In the second stage, tweets were ranked in descending order of model score and the top $p$\% were reviewed by human moderators. The share $p$ was endogenously determined by operational parameters, including total daily tweet volume in the language, the number of moderators assigned to that language, daily working hours per moderator, reviewing speed, and the number of independent reviewers per tweet.
We adopted conservative assumptions designed to minimize human error and moderator overload. In our baseline specification, each tweet was independently reviewed by three moderators and each moderator reviewed 100 tweets per hour. These assumptions likely overstate moderation effort relative to current operational practice and therefore yield upper-bound estimates of resources needed to achieve a certain moderation rate.
Our primary outcome measure is \textit{avoided hate engagement}, defined as the share of total engagement with hateful tweets that would be prevented through timely moderation, where engagement is measured as total interactions accrued within ten minutes of posting. Tweets were ranked using a combined score that incorporates both predicted hatefulness and early engagement in order to prioritize content with the greatest potential exposure. As a secondary measure, we also computed \textit{coverage}, the share of all hateful tweets that are surfaced for review and, in principle, moderated (\ref{si_human_ai_coverage}). Because exposure is measured in engagement rather than in tweets, avoided hate engagement was estimated on the engagement-weighted sample, which is representative of exposure, whereas coverage, a tweet-level quantity, was estimated on the random sample, which is representative of the tweet population.
To approximate moderation capacity, we used moderator counts reported in Twitter’s October 2023 Digital Services Act transparency filing \citep{X_DSA_Transparency_Report_Nov2023}, which provide the earliest publicly available language level staffing data. Given documented workforce reductions after September 2022, we treated these figures as conservative lower-bound estimates of moderation capacity at the time of posting. Because not all moderation activity concerns hate speech, we estimated the effective number of moderators handling hate-related content by applying the share of manual moderation decisions attributed to hate speech (31\% in 2024 \cite{X_Transparency_Report_2025}) to the reported total moderator counts for each language. This adjusted figure was used as the baseline workforce in our simulations.
To contextualize resource requirements, we estimated the annual financial cost of achieving target levels of avoided hate engagement as a share of Twitter's 2021 annual revenue (USD 5.08 billion \citep{Twitter_10K_2021}), the last full year the platform reported before being taken private in late 2022. Cost calculations assumed an hourly wage of USD 20 for English-language moderation, converted to purchasing-power-equivalent hourly wages for the other languages (\ref{si_cost_analysis}), and continuous platform operation (8 hours per day, 365 days per year), reflecting the fact that content is posted every day of the year.
A formal description of the simulation framework, parameter choices and justification, workforce estimation procedures, additional coverage results, and sensitivity analyses is provided in \ref{si_human_ai_simulation}.

\subsection{Data and Sampling}\label{si_data_sampling}

\subsubsection{Full-platform dataset}

We used the TwitterDay dataset \citep{pfeffer2023just}, comprising all public tweets posted globally during a 24-hour period beginning September 21, 2022. Tweets were collected approximately ten minutes after posting using the Academic API. The dataset consists of 375 million original tweets and retweets and provides a complete snapshot of public platform activity during that period.

\subsubsection{Sampling}\label{si_sampling}

\paragraph{Unit of analysis}
Sampling was conducted on original tweets only. Retweets were excluded in order to avoid duplication of content and to ensure that each observation represents a unique authored statement.

\paragraph{Linguistic and geographic scope}
We focused on eight widely used platform languages: Arabic, English, French, German, Indonesian, Portuguese, Spanish, and Turkish. 
Language selection was guided by two criteria. First, we aimed to capture substantial global platform coverage across regions. Second, we restricted attention to languages for which at least one publicly available academic hate speech dataset existed at the time of analysis (August 2024), as cataloged by \url{https://hatespeechdata.com/}. This ensured the availability of benchmark detection models for feasibility analysis. Because enforcement and moderation practices may differ across countries even when the primary language is shared, we additionally sampled tweets from four English-speaking countries spanning multiple continents: the United States, India, Nigeria, and Kenya. This allows us to examine geographic variation holding language constant.

\paragraph{Language inference}
Tweet language was inferred using the language label provided by the Twitter API for each tweet. This label is generated by Twitter’s internal language detection system. The distribution of languages in the full 24-hour TwitterDay dataset is reported in Table~\ref{tab:language_share_twitterday}.

\input{tables/share_lang_twitterday}

\paragraph{Country inference}
User country was inferred by geocoding the self-reported profile location field using the Google Geocoding API. This approach follows established procedures in prior work \citep{hecht2011tweets}. 
Because not all users provide a profile location, and because not all location strings can be successfully mapped to a country, country inference is available for only approximately 25\% of users in the full dataset. As a result, country-based samples are less representative than language-based samples, which cover all tweets.

\paragraph{Sampling strategy and weighting}
For each language and country, we drew two distinct samples:
\begin{enumerate}
    \item A random sample of 30{,}000 tweets.
    \item An engagement-weighted sample of 15{,}000 tweets, where sampling probability was proportional to total early engagement.
\end{enumerate}

Total engagement was defined as the sum of likes, retweets, replies, and quote tweets measured ten minutes after posting. This short time window corresponds to the data collection procedure of the TwitterDay dataset and serves as a proxy for early exposure.

Random samples were used for prevalence estimation, enforcement analysis, AI standalone performance and moderation coverage estimation in the human-AI moderation setup. Engagement-weighted samples were used to study the relationship between engagement and removal and compute avoided engagement with hate in the human-AI moderation setup.

In total, the sampled dataset comprises 540{,}000 tweets across languages and countries.

\subsection{Annotation}\label{si_annotation}

\subsubsection{Annotation team}
The annotation team consisted of 36 annotators, with three annotators assigned to each language–country group. Annotators were recruited through targeted outreach and screened for native or near-native proficiency in the target language, prior annotation experience, and familiarity with local sociolinguistic context. To capture within-language variation, we aimed to include annotators from different national or regional backgrounds where applicable (e.g., different dialect communities within the same language). The team was diverse in gender, age (primarily 18–39), and educational background (undergraduate to doctoral degrees). Compensation ranged from 5 to 24 US dollars per hour depending on country, education, and experience.

\subsubsection{Annotation framework}
\label{si_annotation_framework}

\paragraph{Policy alignment} Tweets were annotated using a prescriptive framework aligned with Twitter’s Hateful Conduct Policy in effect at the time of data collection \citep{twitter2022hateful}. The annotation protocol operationalizes the platform’s policy definitions in a structured coding scheme designed to distinguish hate speech from non-hateful content.

\paragraph{Primary classification} Each tweet was assigned to one of three mutually exclusive categories:

\begin{enumerate}
    \item \textbf{Hate:} Content that promotes or incites violence against, or directly attacks, a protected group, defined by race, ethnicity, nationality, religion, gender, or sexual orientation.
    \item \textbf{Offensive:} Abusive, profane, or insulting language that does not target a protected group.
    \item \textbf{Neutral:} All remaining content.
\end{enumerate}

\paragraph{Violence} Tweets labeled as hate were further categorized as \textit{violent} or \textit{non-violent}. Violent hate includes explicit calls for physical harm, extermination, or direct threats against protected groups. Non-violent hate includes degrading, dehumanizing, or exclusionary attacks (e.g., calls to ban or remove a protected group from public life) that do not contain explicit calls for violence.

\paragraph{Additional content categories} On a subset of 300,000 tweets, corresponding to one third of random samples and all engagement-weighted samples, annotators additionally labeled the presence of other policy-relevant content types to enable comparison of hate speech enforcement with enforcement of other sensitive categories. Specifically, they annotated:

\begin{itemize}
    \item \textbf{Scam:} Content that attempts to deceive users for financial or personal gain, including fraudulent offers, impersonation, or phishing attempts.
    \item \textbf{Adult content:} Sexually explicit, pornographic, or strongly suggestive material.
\end{itemize}

\paragraph{Annotation design}
All tweets were independently labeled by three native speakers of the respective language for the primary three-way classification (hate, offensive, neutral) and for the additional content categories. Tweets labeled as hate were subsequently classified as violent or non-violent by two independent annotators. Annotations were conducted on textual content only. The full annotator-facing codebook, including detailed decision rules and examples, is provided as \texttt{annotation/codebook.md} in the deposited replication materials \citep{tonneau2026replication}.

\paragraph{Attention checks} To ensure annotation quality, we inserted attention checks into each annotation chunk of size 2,500. For each chunk, four positive examples were randomly sampled: two from existing hate speech datasets for the primary annotation task \cite{tonneau-etal-2024-languages} and two from the auxiliary categories generated using GPT-5.1. These items were manually verified by the first author prior to insertion and placed at random ranks within the chunk. After completing each chunk, the first author reviewed the annotator’s responses to the attention check items. If the annotator failed to correctly identify the inserted positives, the entire chunk was returned for re-annotation. This process was repeated until all attention checks in the chunk were correctly passed.

\subsubsection{Inter-annotator agreement}
\label{si_inter_annotator_agreement}

\paragraph{Primary three-way annotation}
Agreement for the primary three-way annotation (hate, offensive, neutral) was assessed using Fleiss’ $\kappa$. Across all samples combined, overall agreement was moderate ($\kappa = 0.50$). 

Across languages, the average $\kappa$ was 0.47 (range: 0.36–0.65). Across English-speaking country samples, the average $\kappa$ was 0.57 (range: 0.50–0.64). Agreement was therefore moderate and broadly consistent across contexts.

Agreement by language and country is reported in Table~\ref{tab:kappa_3class}. 

\input{tables/kappa_3class}

\paragraph{Violent vs.\ non-violent hate}
Agreement for the secondary violent vs.\ non-violent hate classification was assessed using Cohen’s $\kappa$ (Table~\ref{tab:kappa_violence}). Across languages, average agreement was substantial ($\kappa = 0.67$, range: 0.45–0.79). Across English-speaking country samples, the average $\kappa$ was moderate ($\kappa = 0.49$, range: 0.39–0.61). 

\input{tables/kappa_violence}

\paragraph{Additional content categories}
Agreement for the additional content categories (scam, and adult content) was assessed using Fleiss’ $\kappa$ (Table~\ref{tab:kappa_additional}).

Across languages, average agreement was 0.25 for scam and 0.71 for adult content. Across English-speaking country samples, the corresponding averages were 0.33 (scam) and 0.53 (adult content). Scam is both rare and harder to label consistently than adult content, which lowers its $\kappa$ even though raw agreement exceeded 97\% for both. To confirm that the main-text hate-vs.-scam removal contrast does not depend on how scam is labeled, we re-estimate it in \ref{si_scam_adult_agreement} with the scam and adult-content indicators defined at three annotator-agreement thresholds, ranging from a tweet flagged by a single annotator to one flagged by all three. The scam removal premium stays large and significant throughout.

\input{tables/kappa_additional}

\subsubsection{Adjudication procedure}

Tweets with no majority label for primary classification and violence classification were discussed among annotators to reach a final label. 
Tweets with at least one hate label were verified and, where applicable, corrected by the first author.

\subsubsection{Prevalence and composition of hate speech}\label{si_prevalence_composition_stats}
We provide descriptive statistics on the prevalence and composition of hate speech in our annotated representative sets in Figs.~\ref{fig:prevalence_composition_lang} and~\ref{fig:prevalence_composition_country}. While the composition of hate speech varies across languages and countries, the most common categories relate to race, ethnicity, or national origin, followed by gender.

\input{figures/prevalence_composition_stats}

\subsubsection{Exposure and platform-scale estimation}\label{si_exposure_estimation}

\paragraph{Platform-scale extrapolation}

Let $\hat{p}_\ell$ denote the estimated prevalence of hateful tweets in the unweighted random sample for language $\ell$, and let $T_\ell$ denote the total number of original tweets posted in that language during the same 24-hour snapshot. We estimated the total number of hateful tweets in language $\ell$ as

\[
\hat{H}_\ell = \hat{p}_\ell \times T_\ell .
\]

The platform-wide total across all included languages is then

\[
\hat{H} = \sum_{\ell=1}^{L} \hat{H}_\ell 
= \sum_{\ell=1}^{L} \left( \hat{p}_\ell \times T_\ell \right),
\]

where $L$ denotes the number of languages.

To quantify sampling uncertainty, we bootstrapped within each language (1{,}000 resamples at the tweet level). For each replicate, we resampled tweets into three mutually exclusive categories (violent hateful, non-violent hateful, and non-hateful) and recomputed both overall hateful prevalence and the share of hate that is violent, thereby jointly propagating uncertainty in both quantities. For bootstrap replicate $b \in \{1, \dots, 1000\}$, we computed

\[
\hat{H}^{(b)} = \sum_{\ell=1}^{L} \left( \hat{p}_\ell^{(b)} \times T_\ell \right).
\]

The language-level tweet totals $T_\ell$ refer to the count of original tweets only, measured directly from the TwitterDay snapshot. The 0.29\% prevalence reported in the main text is the unweighted mean of per-language prevalences and therefore differs numerically from the volume-weighted aggregate $\hat{H} / \sum_\ell T_\ell$ (0.23\%), which is dominated by high-volume but lower-prevalence languages such as English. As a result, the platform-scale total of 224{,}963 hateful tweets per day across the eight included languages is substantially smaller than a back-of-the-envelope $0.0029 \times 0.54 \times 375$M would suggest: the 375M figure includes retweets, the 0.29\% is unweighted across languages, and the actual sum applies the language-specific prevalence $\hat{p}_\ell$ to the language-specific original-tweet volume $T_\ell$. The language-level tweet totals $T_\ell$ were treated as fixed and measured without error. We reported the point estimate together with a 95\% percentile bootstrap confidence interval,

\[
\hat{H}_{CI} =
\left[
\hat{H}_{2.5}, \,
\hat{H}_{97.5}
\right],
\]

where $\hat{H}_{2.5}$ and $\hat{H}_{97.5}$ denote the 2.5th and 97.5th percentiles of the bootstrap distribution $\{ \hat{H}^{(b)} \}_{b=1}^{1000}$. These intervals capture sampling uncertainty in the snapshot but do not reflect day-to-day temporal variation in hate volume. The same bootstrap applied to the share of hate that is violent yields an estimated 11{,}071 violent hateful original tweets per day (95\% CI [7{,}657, 15{,}205]).

\paragraph{User exposure approximation}

Let $F$ denote the number of tweets viewed per day under a chronological feed. The average user spends approximately 34 minutes per day on the platform \citep{grabon_twitter_statistics_2025}, which we conservatively approximated to about 200 tweets read per day, corresponding to roughly six tweets per minute, broadly consistent with reported scrolling behavior on short-form social feeds. Expected exposure scales linearly in this choice, so alternative viewing rates rescale the estimates below proportionally.

Because a user's feed is populated overwhelmingly by content in their own language, per-user exposure is governed by the prevalence in that language rather than by the volume-weighted platform aggregate. Accordingly, we use the unweighted mean of per-language Twitter-policy hate prevalences, the same quantity reported in the main text,

\[
\hat{p} = \frac{1}{L}\sum_{\ell=1}^{L} \hat{p}_\ell .
\]

Expected daily exposure to hateful tweets is given by

\[
E[\text{hateful tweets per day}] = \hat{p} \times F.
\]

Using $F = 200$, we obtain

\[
E[\text{hate}] = 0.0029 \times 200 = 0.59,
\]

\[
E[\text{violent hate}] = 0.00025 \times 200 = 0.05,
\]

that is, roughly one hateful tweet every two days and one violent hateful tweet about every three weeks. These calculations assume chronological ranking and within-language exposure proportional to prevalence.

\paragraph{Per-language exposure under a monolingual feed} The pooled figure above averages across language communities and no individual user experiences it directly: because most users consume content predominantly in one language, their exposure is set by that language's prevalence. Table~\ref{tab_si:per_language_exposure} reports the expected exposure of a monolingual user, $E_\ell = \hat{p}_\ell \times F$ with $F=200$, using the Twitter-policy hate prevalence $\hat{p}_\ell$ in each language's random sample. Expected exposure ranges from about one hateful tweet every five days (Portuguese, Indonesian) to more than one per day (French, Turkish). Because the pooled prevalence weights all languages equally while the platform-scale totals above are volume-weighted, and because high-volume English carries a relatively low prevalence, neither pooled figure is representative of users in the higher-prevalence communities, who are exposed to substantially more hate.

\input{tables/per_language_exposure}

\subsection{Enforcement measurement}\label{si_enforcement_measurement}

\subsubsection{Removal definition}

Enforcement outcomes were measured using Twitter’s Compliance API in February 2023, approximately five months after the original tweets were posted. For each tweet in the annotated sample, we queried its availability and, when applicable, the status of the associated user account.

A tweet was classified as \emph{removed} if it was no longer retrievable via the API at re-query because the posting account had been suspended or because the individual tweet had been deleted. The Compliance API reports non-retrievability reasons at the user level, including \texttt{suspended} and \texttt{deactivated} account status. User-level \texttt{suspended} status is interpreted as a platform-initiated enforcement action that renders all tweets from the account inaccessible. User-level \texttt{deactivated} status reflects a user-controlled action of deactivating one's own account, which also makes tweets inaccessible via the API. Tweets that became inaccessible through voluntary deactivation, or because the account was set to protected (private), are \emph{not} counted as removed. The API does not provide a reason when an individual tweet is unavailable while the account remains otherwise active.

Accordingly, our primary removal measure combines account suspension and individual tweet deletion. Because the API does not distinguish platform-initiated from user-initiated tweet deletion, this measure should be interpreted as an upper bound on platform-initiated enforcement. To provide a more conservative estimate of clearly platform-initiated action, we separately report removal attributable to user suspension only. Suspension-based measures therefore constitute a lower bound on unambiguous platform enforcement.

\subsubsection{Descriptive statistics}

Tables~\ref{tab_si:count_removal_hate}–\ref{tab_si:count_removal_misc} report counts and enforcement outcomes for all content categories used in the regression analyses.

Table~\ref{tab_si:count_removal_hate} summarizes hate-related content, including violent hate, non-violent hate, and overall hate.  
table~\ref{tab_si:count_removal_problematic} reports additional policy-relevant categories, including scam and adult content.  
table~\ref{tab_si:count_removal_misc} presents remaining covariates included in the regression models.

Removal rates were calculated as the proportion of tweets removed at the five-month re-query, following the definition above. For each category, we reported the total number of tweets, the number removed, the removal rate in percent, and 95\% confidence intervals computed using the Wilson score method for binomial proportions \citep{wilson1927probable}, which, unlike the standard normal approximation, remains reliable for the small counts and near-zero rates common in our data. We additionally reported suspension rates to distinguish clearly platform-initiated enforcement from broader tweet inaccessibility.

\input{tables/count_removal_hate}
\input{tables/count_removal_problematic}
\input{tables/count_removal_misc}

Table~\ref{tab_si:tweet_covariates} reports descriptive statistics for tweet-level covariates used in removal regressions, including retweet count, the Twitter ``sensitive'' flag, and the reply indicator. Continuous variables are reported as means with standard deviations in parentheses. Binary variables are reported as percentages.

\input{tables/tweet_level_metrics}

Table~\ref{tab_si:user_covariates} reports descriptive statistics for user-level covariates included in the regression models, including verified account status and account-level activity measures. Continuous variables (followers, following, lifetime tweet count) are reported as medians. Verified status is reported as a percentage. The tweet-count column reflects the total number of statuses posted on a user's account at the time of collection (\texttt{statuses\_count}, lifetime), not the number of tweets posted within the 24-hour snapshot.

\input{tables/user_level_metrics}

To complement tweet-level enforcement descriptives, table~\ref{tab_si:user_suspension_rates} reports user suspension rates by language and country, including suspension rates among users posting hate speech.

\input{tables/user_suspension_rates}

Table~\ref{tab_si:count_removal_hate_by_reason} decomposes the overall hate-tweet removal rate into its two components, removal because the posting account was suspended and removal of the individual tweet without account-level action, and additionally reports the share of tweets inaccessible because the account was voluntarily deactivated, which is not counted as removed under our definition. Suspension is unambiguously platform-initiated. Individual-tweet deletion is ambiguous between platform and user, so the total constitutes an upper bound on platform-initiated enforcement beyond suspension.

\input{tables/count_removal_hate_by_reason}

\subsubsection{Regression specification and tables}\label{si_reg_spec_tables}

We estimated linear probability models (LPMs) to assess associations between hate speech and enforcement outcomes. The primary tweet-level specification is:

\[
\Pr(\text{Removed}_i = 1) =
\alpha +
\beta_1 \text{ViolentHate}_i +
\beta_2 \text{NonViolentHate}_i +
\gamma' \mathbf{X}_i +
\delta_{\ell(i)} +
\varepsilon_i,
\]

where $\mathbf{X}_i$ includes log-transformed retweet count, the Twitter ``sensitive'' flag, reply status, verified author status, and additional policy-relevant content categories (scam, adult content). The term $\delta_{\ell(i)}$ denotes language fixed effects in the language sample and country fixed effects in the country sample, absorbing cross-language and cross-country differences in baseline removal rates. Neutral content serves as the reference category. English (language sample) and the United States (country sample) are the omitted reference levels for the fixed effects.

To examine cross-linguistic and cross-country disparities, we additionally estimated interaction models between hate indicators and language or country indicators, using English and the United States as reference categories, respectively. The corresponding tables report both the main effect of hate for the reference group (English or US) and the interaction terms capturing differences relative to that reference. All language-level and country-level models were estimated separately.

We further estimated analogous models at the user level, predicting account suspension as a function of whether a user posted hate speech, controlling for user-level characteristics including tweet volume, follower count, following count, and verified status, plus language or country fixed effects.

To assess potential visibility filtering, we estimated linear models predicting log-transformed total engagement as a function of hate indicators and the same control variables (excluding retweet count, which is mechanically a component of engagement), together with language or country fixed effects. Engagement is defined as $\log(1 + \text{likes} + \text{replies} + \text{retweets} + \text{quotes})$.

Standard errors are plain OLS for the main tweet-level and user-level models, and heteroskedasticity-robust (HC1) for the interaction models reported below. Robustness to user-level clustering is reported in \ref{si_robustness_checks}.

\paragraph{Tweet-level removal}  
Tables~\ref{tab_si:tweet_removal_language_reg} and~\ref{tab_si:tweet_removal_country_reg} report removal regressions for the language and country samples, respectively. Tables~\ref{tab_si:language_interaction_removal} and~\ref{tab_si:country_interaction_removal} present the corresponding interaction models. To test whether the hate effect varies across contexts, we estimate $\textit{removed} \sim \textit{hate} \times C(\textit{unit})$ with the full column-(4) controls and heteroskedasticity-robust (HC1) standard errors, and jointly test all hate$\times$unit interaction terms. The hate effect is not significantly different across languages (joint Wald $F_{7}=1.52$, $p=0.15$) and only marginally so across countries ($F_{3}=2.13$, $p=0.09$). Because these are non-significant tests on small samples, we report below the equivalence bounds and statistical power that quantify what they can and cannot rule out (table~\ref{tab_si:removal_equivalence_power}).

\input{tables/tweet_level_removal_reg_lang_claude}
\input{tables/tweet_level_removal_reg_country_claude}
\input{tables/tweet_level_removal_reg_interaction_lang_claude}
\input{tables/tweet_level_removal_reg_interaction_country_claude}

\paragraph{Equivalence bounds and statistical power for the removal nulls}
Several of our removal estimates are statistical nulls, and all of them rest on modest numbers of hateful tweets (706 in the language sample, 329 in the country sample, and as few as 60 violent hate tweets), so a non-significant result cannot by itself be read as evidence of no effect. We therefore report each estimate with its verified 95\% confidence interval and the minimum detectable effect (MDE), the smallest true premium the test would detect in 80\% of repeated samples, and, for the nulls, a two one-sided tests (TOST) equivalence assessment, all computed with the HC1 robust standard errors used above (table~\ref{tab_si:removal_equivalence_power}). Equivalence testing does not apply to the pooled country premium, which is significantly positive, or to the joint interaction tests, which test all interaction terms at once and therefore yield no single effect size to bound, so those rows report confidence intervals and MDEs only. For the pooled hate premium, we pre-specify an equivalence bound of $\Delta=\pm5$ percentage points, the largest premium we would still treat as practically negligible. The comparison of interest is whether hate is actioned like the other policy categories we benchmark against, which carry premiums of $+29$pp (scam) and $+17$pp (adult content). A premium below $\Delta$ is at most about one sixth of the scam premium and under a third of the adult-content premium, and so would not represent prioritization on the scale of those categories.

\input{tables/removal_equivalence_power}

The pooled hate premium is bounded within a few percentage points. In the language sample, it is $+2.6$pp (95\% CI $[-0.4,+5.6]$), and the equivalence test rejects premiums larger than $\pm5.2$pp. This is marginally wider than our pre-specified bound, so equivalence at $\Delta=\pm5$pp is not formally established, though the data remain clearly inconsistent with a premium approaching those of scam and adult content. The estimate is also sensitive to specification. With tweet-level controls but no language fixed effects, it is $+3.3$pp ($p=0.03$), attenuating to $+1.8$pp and losing significance once the full controls and language fixed effects are added ($p=0.24$, table~\ref{tab_si:tweet_removal_language_reg}). In the country sample, the premium is larger and significantly positive ($+5.5$pp, 95\% CI $[+1.6,+9.5]$), but it is driven by India: excluding the Indian sample under the same specification, it falls to $+1.3$pp and is no longer significant ($p=0.55$).

We do not claim equivalence for the cross-language and country interactions or for the violent hate comparison, because these tests are markedly underpowered. With 30 to 180 hateful tweets per language, the hate$\times$language contrasts detect only differences of roughly 18 to 26pp, the hate$\times$country contrasts 16 to 18pp, and the violent versus non-violent comparison about 15pp on 60 violent hate tweets. The estimates are correspondingly noisy: per-context premiums range from about $-8$pp (Arabic) to $+14$pp (Indonesian), with no individual contrast surviving correction for multiple testing, and violent hate is removed no more often than non-violent hate under the full controls ($-1.8$pp, $p=0.76$). These tests therefore rule out only differences larger than roughly 15 to 26pp, depending on the comparison, and cannot exclude differences as small as the pooled premium itself.

Finally, the comparison with scam and adult content in the main text rests on a single full-control specification that enters aggregated hate alongside the two other categories. It estimates the hate coefficient at approximately zero ($+0.2$pp, $p=0.93$, $n=80{,}000$) while recovering the $+29$pp scam and $+17$pp adult premiums. We report this coefficient here rather than as a separate column of table~\ref{tab_si:tweet_removal_language_reg}, which instead splits hate into its violent and non-violent components.

\paragraph{User-level suspension}
Tables~\ref{tab_si:user_suspension_language_reg} and~\ref{tab_si:user_suspension_country_reg} report user-level suspension models for the language and country samples, and Tables~\ref{tab_si:language_interaction_suspension} and~\ref{tab_si:country_interaction_suspension} the corresponding interaction specifications. As with removal, a joint Wald test rejects equality of the hate effect across units (language $F_{7}=3.70$, $p<0.001$; country $F_{3}=3.75$, $p=0.010$). Only the Arabic contrast is individually significant ($p=0.046$), however, and no individual hate$\times$language or hate$\times$country interaction survives Bonferroni or Benjamini--Hochberg correction within its family. The cross-context variation in suspension is therefore diffuse and, like the removal interactions, cannot be attributed to any single language or country.

\input{tables/user_level_suspension_reg_language}
\input{tables/user_level_suspension_reg_country}


\input{tables/user_level_suspension_reg_interaction_lang}
\input{tables/user_level_suspension_reg_interaction_country}

\paragraph{Engagement}
Tables~\ref{tab_si:engagement_language} and~\ref{tab_si:engagement_country} report linear models predicting log total engagement from hate indicators and controls, estimated on the engagement-weighted samples. These test whether the platform systematically reduced the visibility of hateful content, which would show up as lower engagement for hateful tweets than for otherwise comparable ones. Violent hate is not negatively associated with engagement in either sample (language $+0.62$, $p=0.11$; country $+0.32$, $p=0.45$), so we find no evidence that the most severe hateful content was systematically downranked. Non-violent hate carries a small negative coefficient in the language sample ($-0.21$, $p=0.005$), but this is not reproduced in the country sample ($+0.07$, $p=0.48$).

Table~\ref{tab_si:engagement_removal_interaction} reports a separate interaction model predicting tweet removal as a function of hate, log total engagement, and their interaction. Fig.~\ref{fig:low_enforcement}C plots the implied marginal effect.

\input{tables/engagement_reg_lang_claude}
\input{tables/engagement_reg_country_claude}
\input{tables/engagement_removal_interaction_claude}

\subsubsection{Robustness checks}\label{si_robustness_checks}

We conducted several robustness analyses:

\begin{itemize}
    \item Logistic regression models with identical covariates. Average marginal effects are reported (Tables~\ref{tab_si:robustness_1}–\ref{tab_si:robustness_2}).
    \item Standard errors clustered at the user level to account for within-user correlation (Tables~\ref{tab_si:robustness_clustered_lang}–\ref{tab_si:robustness_clustered_country}).
    \item Sensitivity of the scam and adult-content comparison to the annotator-agreement threshold used to construct each indicator (table~\ref{tab_si:scam_adult_agreement_sensitivity}).
    \item Decomposition of the scam and adult-content removal advantage by enforcement channel (account-level suspension vs.\ tweet-level deletion) and robustness to bot-like account characteristics (table~\ref{tab_si:scam_adult_channel_decomp}).
\end{itemize}

\paragraph{Logistic regression}

Tables~\ref{tab_si:robustness_1} and~\ref{tab_si:robustness_2} replicate the main tweet-level removal specifications using logistic regression, retaining the language and country fixed effects. We report average marginal effects (AMEs), computed as the sample average of individual-level partial effects, to facilitate direct comparison with the linear probability model estimates. The AME estimates qualitatively track the LPM results across all reported columns of Tables~\ref{tab_si:tweet_removal_language_reg} and~\ref{tab_si:tweet_removal_country_reg}: the aggregate hate indicator is positively associated with removal (significant in the country sample, marginal in the language sample), and this aggregate effect is driven by non-violent hate. The violent hate coefficient alone is not significantly different from zero in either sample.

Column~(4) (with scam and adult content controls) is reported under logit for the language sample but omitted for the country sample, where the policy-annotated subsample contains only 8 violent hate tweets, causing the logit to fail to converge for the violent hate coefficient. The close agreement between LPM and logit estimates in the language-sample column~(4) and in columns (1)--(3) of both samples suggests no estimator-driven difference in the country-sample column~(4).

\input{tables/logistic_ame_lang_claude}
\input{tables/logistic_ame_country_claude}

\paragraph{User-clustered standard errors}

Tables~\ref{tab_si:robustness_clustered_lang} and~\ref{tab_si:robustness_clustered_country} replicate the main tweet-level removal specifications with standard errors clustered at the user level, to account for potential within-user correlation across tweets. Of the 240,000 tweets in the language sample, 214,624 come from unique users, indicating moderate repeat-posting. The patterns from the main LPM tables are robust to user-level clustering: the aggregate hate coefficient remains positive (significant in the country sample, marginal in the language sample), driven by non-violent hate, while the violent hate coefficient alone remains not significantly different from zero. Column~(4) is reported for the language sample but omitted for the country sample, where the policy-annotated subsample contains only 8 violent hate tweets and the cluster-robust standard error estimator becomes numerically unstable.

\input{tables/clustered_se_lang_claude}
\input{tables/clustered_se_country_claude}

\paragraph{Removal and annotator agreement}\label{si_annotator_agreement_removal}

We test whether enforcement is stronger for tweets whose hatefulness is less ambiguous, that is, those independently flagged as hate by more annotators. Among hateful tweets, we group each tweet by the number of annotators out of three who labeled it as hate and compare removal across these agreement levels. Removal rates are essentially flat: $21.3\%$ for tweets flagged by one annotator ($N=169$), $22.5\%$ for two ($N=151$), and $20.4\%$ for all three ($N=103$). Estimating a linear probability model of removal on indicators for two- and three-annotator agreement, with the full column-(4) controls, language fixed effects, and author-clustered standard errors, we find no differentiation by agreement: a joint test that removal is unrelated to agreement level is far from significant ($F_{2}=0.20$, $p=0.82$), and unanimously flagged hate is not removed at a higher rate than hate flagged by a single annotator ($\hat{\beta}=+2.5$ percentage points, $p=0.66$). Enforcement is thus not concentrated on the clearest-cut cases of hate.

\paragraph{Annotator-agreement threshold for the scam and adult-content comparison}\label{si_scam_adult_agreement}

The hate-vs.-scam and hate-vs.-adult-content comparisons are rhetorically central: the main text reads the column~(4) coefficients as showing that other policy categories carry substantial removal effects while hate does not. Adult content rests on firm inter-annotator agreement (Fleiss' $\kappa = 0.71$ across languages), but scam annotation is noticeably noisier (overall $\kappa = 0.25$, with cross-language values as low as $\kappa = 0.13$ in Indonesian and $\kappa = 0.15$ in the Nigerian sample). To verify that the hate/scam contrast is not an artifact of the headline majority-vote ($\geq$2/3) threshold, we refit column~(4) of the language-sample removal regression under three alternative annotator-agreement thresholds for each indicator: at least one annotator flagging the tweet ($\geq$1/3), majority ($\geq$2/3, the headline spec), and unanimous (3/3).

Table~\ref{tab_si:scam_adult_agreement_sensitivity} reports the resulting scam and adult coefficients, with the hate coefficients shown for reference. At the headline majority threshold, the scam coefficient is $\hat{\beta}=0.29$, reproducing the column~(4) estimate. Restricting to unanimous (3/3) flags shrinks the scam-positive set from 148 to 34 tweets and the coefficient remains positive and significant ($\hat{\beta}=0.16$, $p<0.01$), while at the loosest $\geq$1/3 threshold it is $\hat{\beta}=0.10$ ($p<0.001$). In every cell, the coefficient remains substantially larger than the violent and non-violent hate coefficients, which stay close to zero and not significantly different from it across all thresholds. The adult coefficient, anchored on much higher agreement, increases under stricter agreement ($\hat{\beta}=0.13$, $0.17$, $0.21$ at the $\geq$1/3, $\geq$2/3, and 3/3 thresholds). The headline contrast, namely that hate is not removed at higher rates than neutral content, while scam and adult content are, is therefore preserved under any agreement criterion, and is supported most cleanly by the adult-content comparison.

\input{tables/scam_adult_agreement_sensitivity}

\paragraph{Enforcement channel and bot-like account characteristics}\label{si_scam_adult_channel}

Because scam and adult-content accounts may be disproportionately automated, a natural concern is that the scam and adult removal advantage over hate reflects platform-initiated bot sweeps rather than content-policy prioritization. We probe this in two ways, holding the column~(4) specification of table~\ref{tab_si:tweet_removal_language_reg} fixed (table~\ref{tab_si:scam_adult_channel_decomp}). First, we decompose removal into its mutually exclusive channels by re-estimating the specification with the dependent variable swapped for each removal reason. Both advantages are realized almost entirely through account-level \emph{suspension} (scam $+35$ percentage points, adult $+17$ percentage points, both $p<0.001$) rather than tweet-level \emph{deletion}, where neither category carries a positive premium (adult $+0.3$ pp, $p=0.70$, while the scam-deletion cell contains only five tweets and is too thin to interpret). Second, this account-level concentration does not make the contrast a bot-sweep artifact: adding bot-like account signals (the log follower-to-following ratio and log lifetime tweets per day) on top of the user-level controls already in column~(4) attenuates the raw advantage by only about a quarter (scam from $+38$ to $+27.5$ pp, adult from $+23$ to $+17$ pp) and leaves a large, highly significant residual. Finally, for adult content, the advantage is not attributable to image-based sensitive-media detectability: only 119 of the 1{,}269 adult tweets carry the platform's sensitive-media flag, and the removal premium is, if anything, larger among text-only adult tweets ($+17.6$ pp) than among those with flagged media ($+11.1$ pp). Throughout, the hate coefficients remain statistically indistinguishable from zero, and every conclusion is unchanged under user-clustered standard errors.

\input{tables/scam_adult_channel_decomp}

\paragraph{Changing leadership and reinstated accounts}

Tweets in our dataset were posted in September 2022, before Elon Musk's acquisition of Twitter in late October 2022, while enforcement was measured five months later in February 2023. Because the pre-acquisition moderation team remained in place for approximately one month before the ownership change, and because the subsequent leadership reduced moderation capacity and relaxed enforcement rules, most enforcement actions were likely taken before the transition. However, one concern is that the post-acquisition leadership reinstated a number of previously suspended accounts \citep{travisbrown_reversals}, which could reduce observed removal rates if tweets from formerly suspended authors became accessible again.

To assess this, we cross-referenced all author IDs in our dataset against a list of accounts known to have had their suspensions reversed following the acquisition \citep{travisbrown_reversals}. We identified 31 overlapping authors (0.008\% of the 399,470 unique authors in our data), responsible for 69 tweets, none of which were annotated as hateful. We re-estimated all regression models, tweet-level removal (language and country, random samples), engagement (language and country, engagement-weighted samples), and the engagement$\times$removal interaction, after excluding these 69 tweets. Estimates, standard errors, and significance levels are identical to the full-sample results across all models. Account reinstatements therefore pose no threat to our conclusions.

\paragraph{Keyword search for pre-collection enforcement}\label{si_kw_search}

Because tweets were collected approximately ten minutes after posting, we cannot rule out enforcement occurring within this window, before collection. To assess the likelihood of immediate pre-collection moderation, we conducted a keyword search for nine highly explicit violent hate terms across the full TwitterDay dataset. The list was constructed to surface only \emph{unambiguous} violent hate (direct calls to kill or exterminate a protected group), covering the four most-spoken languages in our sample for which we could compile a defensible translation. It is not intended to be exhaustive but serves as a worst-case probe: because such unambiguous terms are the content most easily caught by keyword filtering, their presence in our collected data would indicate that even the most trivially detectable hate was not removed within the first ten minutes, pointing to minimal enforcement in this time period. The keywords were: \textit{kill all jews}, \textit{gas the jews}, \textit{death to jews}, \textit{kill all n****rs}, \textit{n****rs must die}, \textit{exterminate muslims}, \textit{mort aux juifs} (French: ``death to Jews''), \textit{juden vergasen} (German: ``gas the Jews''), and \textit{muerte a los judios} (Spanish: ``death to Jews''). We retrieved 90 matches in total.

After qualitative author review of each hit, 5 were true positives. Examples include a tweet reading ``f**k n****rs kill all n****rs'', a Spanish-language tweet reading ``hitler es god, muerte a los judios, putos negros'', and an English-language tweet invoking religious eschatology to call approvingly for the killing of Jews.

These cases demonstrate that explicit, unambiguous violent hate was present in the data at the time of collection. We cannot rule out some pre-collection enforcement, but the survival of such explicit violations points to minimal enforcement within the first ten minutes. The keyword list and search script are included in the deposited replication materials \citep{tonneau2026replication}.

\paragraph{Suspension attribution: is suspension driven by the annotated content?}\label{si_suspension_attribution}

A natural concern with using account suspension as a moderation outcome is that Twitter may have suspended an account for reasons unrelated to the specific tweet we annotated. If suspension reflects only a user's broader posting pattern, for example, repeated scam content posted elsewhere, then the per-tweet labels we use would simply be proxies for ``users who get suspended for unrelated reasons,'' and our conclusions about the relationship between hate content and moderation would be confounded.

To assess this, we collected every tweet posted by the annotated users during the 24-hour TwitterDay window (2022-09-20 15:00 UTC to 2022-09-21 14:59 UTC) and asked whether the user's \emph{other} (non-annotated) tweets independently predict their suspension status. We restrict this analysis to users in the eight language buckets, for which we score hate on other tweets using the per-language Perspective \textsc{Identity\_Attack} threshold derived from the HateDay benchmark (German tweets scored with \texttt{jagoldz-gahd}). Adult and scam flags are obtained from lexical/regex classifiers tuned to the dataset. We consider three sets of users: \emph{Group A}, users whose annotated tweet was labeled as scam or adult content ($n=3{,}770$); \emph{Group B}, users whose annotated tweet was labeled as Twitter-defined hate ($n=902$); and \emph{Group C}, a control set of users whose annotated tweet was labeled as neutral, stratified-matched to the union of A and B on language ($n=4{,}602$). Table~\ref{tab_si:attribution_descriptive} reports descriptive statistics by annotation group and suspension status.

\input{tables/si_attribution_descriptive}

Two patterns motivate the regression that follows. Across all groups, suspended users fit the generic ``bad-account'' profile of recently created, low-reputation accounts and post more flagged content elsewhere than their non-suspended counterparts, consistent with the concern that suspension is partly driven by behavior beyond the annotated tweet. This is starkest in the neutral control group: Group~C-suspended users post a strikingly high share of scam content in their non-annotated tweets (23.2\%, versus 0.2\% for their non-suspended counterparts) and have a bot-like profile (median age 112 days, 21 followers, 20 following), indicating that a small fraction of ``neutral'' suspensions reflect platform action against spammers whose single annotated tweet happened to land on inoffensive content.

To test whether the annotation labels carry signal \emph{beyond} this generic ``bad-account'' explanation, we fit logistic regressions of suspension on (1) the share of non-annotated tweets flagged as hate, adult, and scam; (2) annotation-group dummies (with Group~C as the reference); (3) user-metadata controls (log non-annotated tweet count, log followers, log following, log tweets per day, verified status); and (4) language fixed effects. Results are reported in table~\ref{tab_si:attribution_regression}.

\input{tables/si_attribution_regression}

Column~(1) confirms that other-tweet content predicts suspension, with strong positive coefficients on shares of hate, adult, and scam content. Adding the annotation-group dummies, account-level controls, and language fixed effects (columns~2--4) leaves Group~A (scam/adult) large and significant throughout, while the smaller Group~B (hate) effect reaches significance once account features and language are absorbed. In our preferred specification (column~4), Group~A is associated with $\hat{\beta}=1.52$ ($p<0.001$, $\approx 4.6\times$ odds ratio) and Group~B with $\hat{\beta}=0.51$ ($p<0.001$, $\approx 1.7\times$ odds ratio) higher suspension odds than neutral controls, conditional on identical other-tweet content shares, user-metadata profiles, and language context.

These results have two implications. First, the concern above is partially warranted: suspension is content-driven beyond the annotated tweet, and a small fraction of suspensions in the neutral-control group are clearly unrelated to the annotated content (the bot-profile Group~C-suspended users). Second, and decisively for our main analysis, the annotation labels carry substantial \emph{independent} predictive power for suspension after controlling for both behavior in other tweets and user-level account characteristics. Were the labels merely a proxy for users who get suspended anyway, the group dummies would collapse toward zero in column~(4). They do not. Because the other-tweet content controls are imperfect proxies (Perspective and lexical classifiers, unlike the human-labeled annotated tweet), this test isolates content-level signal rather than action on the specific tweet. That suffices for our purpose: suspensions track hateful and scam/adult content rather than unrelated platform action, so our use of suspension as a moderation outcome is well-founded.

\subsubsection{Examples of non-removed violent hate}\label{si_enforcement_examples_unremoved_violent}

We provide examples of violent hate that remained online five months after being posted. We translated to English and reformulated to avoid re-identification:
\begin{itemize}
    \item Tell that cockroach Paul Kagame we’ve saved a special spot for your little soldiers, six feet under. Your mother’s corpse, you Tutsi!
    \item These muslim pigs are shifting bulshit everywhere. Hunt these b*stards
    \item Let’s eradicate the indigenous people, starting with you
    \item These Palestinian creatures have filthy speech; they don't deserve sympathy, they deserve to be beaten with sandals
    \item To that Greek fantasist: if a Turkish–Greek war breaks out, find me, I’ll shove the barrel of my gun up your ass.
    \item We love Russians with holes in them.
\end{itemize}

\subsection{AI detection evaluation}
\label{si_ai_detection}

\subsubsection{Model selection}

We evaluated publicly available hate speech detection systems across all study languages and country contexts. For supervised learning approaches, we identified the five most widely used open-source hate speech classifiers per language available on Hugging Face, as determined by cumulative download counts in August 2024. Models were included if they (i) were publicly accessible, (ii) were explicitly designed for hate speech or abusive language detection, and (iii) supported the language under study. If fewer than five eligible models were available for a given language, all identified models were included.

\paragraph{Language-specific models}

\begin{itemize}

\item \textbf{Arabic:} 
\texttt{Hate-speech-CNERG/dehatebert-mono-arabic}; 
\texttt{IbrahimAmin/marbertv2-finetuned-egyptian-hate-speech-detection}.

\item \textbf{English:} 
\texttt{facebook/roberta-hate-speech-dynabench-r4-target}; 
\texttt{Hate-speech-CNERG/bert-base-uncased-hatexplain}; 
\texttt{Hate-speech-CNERG/dehatebert-mono-english}; 
\texttt{IMSyPP/hate\_speech\_en}; 
\texttt{pysentimiento/bertweet-hate-speech}.

\item \textbf{French:} 
\texttt{Hate-speech-CNERG/dehatebert-mono-french}; 
\texttt{Poulpidot/distilcamenbert-french-hate-speech}.

\item \textbf{German:} 
\texttt{deepset/bert-base-german-cased-hatespeech-GermEval18Coarse}; 
\texttt{Hate-speech-CNERG/dehatebert-mono-german}; 
\texttt{jagoldz/gahd}; 
\texttt{jorgeortizv/BERT-hateSpeechRecognition-German}; 
\texttt{shahrukhx01/gbert-hasoc-german-2019}.

\item \textbf{Indonesian:} 
\texttt{Hate-speech-CNERG/dehatebert-mono-indonesian}.

\item \textbf{Portuguese:} 
\texttt{Hate-speech-CNERG/dehatebert-mono-portugese}.

\item \textbf{Spanish:} 
\texttt{Hate-speech-CNERG/dehatebert-mono-spanish}; 
\texttt{pysentimiento/robertuito-hate-speech}.

\item \textbf{Turkish:} 
\texttt{ctoraman/hate-speech-berturk}.

\end{itemize}

\paragraph{Country-specific models}

\begin{itemize}

\item \textbf{India:} 
\texttt{Hate-speech-CNERG/bengali-abusive-MuRIL}; 
\texttt{Hate-speech-CNERG/english-abusive-MuRIL}; 
\texttt{Hate-speech-CNERG/kannada-codemixed-abusive-MuRIL}; 
\texttt{Hate-speech-CNERG/marathi-codemixed-abusive-MuRIL}; 
\texttt{Hate-speech-CNERG/tamil-codemixed-abusive-MuRIL}.

\item \textbf{Kenya:} No model was found for Kenya. 

\item \textbf{Nigeria:} 
\texttt{worldbank/naija-xlm-twitter-base-hate}.

\item \textbf{United States:} 
\texttt{facebook/roberta-hate-speech-dynabench-r4-target}; 
\texttt{Hate-speech-CNERG/bert-base-uncased-hatexplain}; 
\texttt{Hate-speech-CNERG/dehatebert-mono-english}; 
\texttt{IMSyPP/hate\_speech\_en}; 
\texttt{pysentimiento/bertweet-hate-speech}.

\end{itemize}

We additionally evaluated Perspective API as a widely deployed commercial toxicity detection system (using Identity Attack as closest to our hate speech definition) and GPT-5.1 as a state-of-the-art decoder-based large language model in a zero-shot setting.

\subsubsection{Evaluation setup}

\paragraph{Classification performance} All models were evaluated fully out-of-sample on our annotated unweighted representative datasets. No additional fine-tuning, retraining, or parameter adjustment was performed. Also, no additional preprocessing beyond standard tokenization required by each model was applied. Supervised models were used with their publicly released pre-trained weights. 

GPT-5.1 was evaluated in a zero-shot setting using the standardized classification prompt below. Temperature was set to 1.0, and each tweet was classified independently. Because GPT-5.1 produces discrete class predictions rather than continuous scores, we reported precision and recall rather than AP.

\begin{quote}
\small\ttfamily
You are an expert in hate speech detection.\\
Classify the following social media post as either \textbf{hate (1)} or \textbf{non-hate (0)} according to the definition and examples below.\\[0.5em]
\textbf{Definition of hate speech:}\\
Hate speech is language that promotes violence or directly attacks, abuses, or threatens an individual or group based on identity characteristics such as race, ethnicity, national origin, caste, sexual orientation, gender, gender identity, religion, age, disability, or serious disease.\\[0.5em]
\textbf{Hate speech often includes:}\\
- Threatening language (e.g., calling for harm against a group)\\
- Dehumanization (e.g., describing a group as animals, insects, or trash)\\
- Support for hateful entities (e.g., denying genocides, praising hateful leaders)\\
- Derogation (e.g., insults, slurs, or demonization of a group)\\
- Animosity (e.g., subtle or implicit negativity toward a group based on identity)\\[0.5em]
If the post does \textbf{not} target an identity group, or is merely offensive, political, or uses general profanity, classify it as \textbf{non-hate (0)}.\\[0.5em]
Return only a single number:\\
- 1 → hate\\
- 0 → non-hate\\[0.5em]
Post: \{tweet text\}
\end{quote}

For supervised models and Perspective API, we computed average precision (AP) based on the continuous hatefulness scores returned by each model. Average precision was chosen as the primary evaluation metric because it is robust to class imbalance and reflects ranking quality across decision thresholds. For each language or country context, we selected the supervised model with the highest average precision (AP) on our representative test set for subsequent analyses. Table~\ref{tab_si:best_models_ap} reports the best-performing model and corresponding AP for each language and country context. Perspective API (Identity Attack attribute) achieved the highest AP in all contexts except the United States, where \texttt{Hate-speech-CNERG/dehatebert-mono-english} performed best.

\input{tables/best_models_ap_claude}

\paragraph{Ranking performance} To assess ranking behavior, we computed the percentile rank of each hateful tweet within the full score distribution for its language. Percentiles were calculated as the proportion of tweets assigned lower or equal hatefulness scores. Median percentile ranks reported in the main text therefore indicate the share of content ranked above a given hateful tweet by the model.

\subsubsection{Error analysis}\label{si_ai_error_analysis}

The analysis described in the main text was conducted through author inspection of non-hateful tweets in the top 0.5\% of the hatefulness score distribution for each language and country context. Offensive content, which uses profane or aggressive language without targeting a protected group, is the single largest source of these false positives, accounting for 38\% of the top 0.5\% of tweets by hate score on average across languages. The remainder consists mainly of tweets that quote or mention hate rather than express it, and cases whose classification depends on context or spelling. Full details, per-language breakdowns, and illustrative examples are provided in \citet{tonneau-etal-2025-hateday}.

\subsection{Human-AI moderation simulation}\label{si_human_ai_simulation}

\subsubsection{Formal framework of the human--AI moderation simulation}
\label{si_human_ai_formal}

We model a two-stage human--AI moderation pipeline operating at the language level. 
For each language $l$, let:

\begin{itemize}
    \item $V_l$ denote the total number of tweets posted per day,
    \item $H_l$ denote the number of hateful tweets per day,
    \item $M_l$ denote the number of moderators assigned,
    \item $r$ denote the review rate (tweets reviewed per moderator per hour),
    \item $h$ denote the number of working hours per moderator per day,
    \item $k$ denote the number of independent reviewers per tweet.
\end{itemize}

The total number of tweets that can be reviewed per day in language $l$ is:

\[
N^{\text{review}}_l = \frac{M_l \cdot r \cdot h}{k}.
\]

Given daily tweet volume $V_l$, the fraction of tweets that can be reviewed is:

\[
p_l = \frac{N^{\text{review}}_l}{V_l}.
\]

Tweets are ranked in descending order of predicted hatefulness score $\hat{h}_i$. For engagement reduction simulations, tweets are ranked by a combined score:

\[
S_i = \hat{h}_i \times \log(1 + e_i),
\]

where $\hat{h}_i$ denotes the predicted hatefulness score and $e_i$ the early engagement of tweet $i$. Engagement enters in log form so that a small number of viral outliers do not dominate the ranking and crowd out lower-engagement but higher-hatefulness content. The hatefulness score is already on a bounded $[0,1]$ scale and is left untransformed.

The top $p_l$ fraction of tweets are selected for human review.

\paragraph{Coverage}
Coverage is defined as the proportion of all hateful tweets that are surfaced for review:

\[
\text{Coverage}_l = 
\frac{\# \{ i \in H_l : i \text{ ranked in top } p_l \}}{H_l}.
\]

\paragraph{Avoided hate engagement}
Let $e_i$ denote early engagement (measured within ten minutes of posting). 
Engagement reduction is defined as:

\[
\text{Engagement Reduction}_l =
\frac{\sum_{i \in H_l^{\text{reviewed}}} e_i}
     {\sum_{i \in H_l} e_i},
\]

where $H_l^{\text{reviewed}}$ denotes the subset of hateful tweets selected for review.

Under the assumption that reviewed hateful tweets are successfully moderated prior to further exposure, this quantity represents the share of hate-related engagement that would be prevented through moderation.

\subsubsection{Parameter justification}\label{si_human_ai_parameters}

We detail below the choice of each operational parameter. Sensitivity of the main results to these assumptions is examined in \ref{si_human_ai_sensitivity}.

\paragraph{Review rate} We assumed each moderator reviewed 100 tweets per hour. \citet{barrett2020moderates} reports that platform moderators process 600–800 items per day, implying roughly 75–100 items per hour over an 8-hour shift. Because our simulation focused exclusively on short text-only content (median tweet length is well under 280 characters), which is faster to evaluate than mixed text--image or text--video items typical of \citet{barrett2020moderates}, we adopted the upper end of this range (100 tweets/hour). The sensitivity of our cost conclusions to this assumption is reported in table~\ref{tab_si:sensitivity_params}.

\paragraph{Reviewers per tweet} We assumed three independent reviewers per tweet. This conservative choice was motivated by the moderate inter-annotator agreement we observed in our annotations (Fleiss' $\kappa = 0.50$), consistent with the well-documented ambiguity at the hate/offensive boundary. This assumption is conservative relative to current practice at large platforms, where many moderation decisions are taken by a single reviewer or algorithmically. Sensitivity to a one-reviewer-per-tweet alternative is reported in table~\ref{tab_si:sensitivity_params}.

\paragraph{Share of workforce dedicated to hate speech} Not all moderation effort concerns hate speech. We estimated the effective share of each language's moderator workforce dedicated to hate-related content by applying the proportion of manual moderation decisions attributed to hate speech reported in Twitter/X's 2024 transparency report (31\%, \citealp{X_Transparency_Report_2025}) to the total per-language moderator counts from the October 2023 DSA filing \citep{X_DSA_Transparency_Report_Nov2023}. This applies a 2024 share to 2023 staffing and assumes the share of moderation effort directed at hate speech was approximately stable across that period. We cannot verify this assumption directly from the available transparency reports.

\subsubsection{Additional results}\label{si_human_ai_coverage}

Under reported staffing (\ref{si_human_ai_parameters}), timely moderation would prevent an estimated 24.5\% of exposure to hate on average across the six languages with reported moderator counts (Fig.~\ref{fig:human-in-the-loop}A), with meaningful reductions for English, French, and German and near-zero reductions for Arabic, Spanish, and Portuguese.

Fig.~\ref{fig:si_coverage_results}A shows moderation coverage as a function of the number of hate-speech moderators, separately for each of the eight language communities. Coverage under current staffing levels (dashed vertical lines) was low across all languages, ranging from roughly 1\% (Spanish) to 25\% (German), reflecting the large gap between available workforce and daily content volume. Languages differed substantially in how many moderators were needed to achieve meaningful coverage: Indonesian and Turkish reached high coverage at lower moderator counts, whereas English required moderator counts in the thousands before coverage rose appreciably. This reflects differences in daily tweet volume and model detection performance across languages.

Fig.~\ref{fig:si_coverage_results}B relates coverage to moderation cost for the two regulated languages, English and German, mirroring the exposure-based frontiers in the main text (Fig.~\ref{fig:human-in-the-loop}B). Reaching high coverage of all hateful tweets is far more expensive than achieving the equivalent reduction in exposure. Across all eight languages combined, achieving 80\% coverage would require approximately 16.6\% of annual revenue at the baseline review intensity (table~\ref{tab_si:sensitivity_params}), roughly twice the 7.9\% needed to avoid 80\% of exposure (main text) and above the 10\% ceiling set by the UK Online Safety Act. This gap illustrates why engagement-weighted targeting is a more tractable policy goal than full coverage: because hateful engagement is concentrated in a small share of hateful tweets (\ref{si_engagement_concentration}), prioritizing high-engagement content for review yields large reductions in user exposure at a fraction of the cost required to review all hateful content.

\input{figures/si_coverage_results}

\subsubsection{Concentration of hateful engagement}\label{si_engagement_concentration}

The tractability of engagement-weighted targeting rests on how unequally engagement is distributed across hateful tweets. We quantify this concentration on the random (tweet-representative) sample, pooling the 706 hateful tweets across the eight languages. The random sample is the appropriate frame here because the quantity of interest is a share \emph{of tweets}, and it is the only valid one: engagement-weighted sampling selects tweets with probability proportional to their early engagement, so the two-thirds of hateful tweets with no early engagement have zero selection probability and are absent from that sample entirely. Inverse-probability weighting cannot recover this unsampled mass, so only the uniform random sample represents the full engagement distribution across hateful tweets. Early engagement, measured as total interactions within ten minutes of posting, is extremely concentrated. First, $66.9\%$ of hateful tweets ($95\%$ Wilson CI $[63.3, 70.2]$) receive no early engagement at all, so they generate essentially no exposure. Second, among the remainder engagement is heavily right-skewed: the top $11\%$ of hateful tweets by engagement account for $80\%$ of all hateful engagement, and the most-engaged $1\%$ carries roughly half. The Gini coefficient of engagement across hateful tweets is $0.89$, close to maximal inequality. This concentration is why prioritizing high-engagement content prevents most exposure while surfacing only a small fraction of hateful tweets, so that avoiding $80\%$ of exposure ($7.9\%$ of revenue) is much cheaper than reviewing $80\%$ of hateful tweets ($16.6\%$). The pattern is uniform across languages (table~\ref{tab_si:engagement_concentration}), with per-language Gini coefficients between $0.79$ and $0.93$. Because the extreme tail rests on few tweets (the pooled top $1\%$ corresponds to roughly seven tweets), the exact top-$1\%$ share is imprecise. The ``top $11\%$ carry $80\%$'' summary and the Gini are the more robust statistics.

\input{tables/engagement_concentration}

\subsubsection{Sensitivity analysis}\label{si_human_ai_sensitivity}

Table~\ref{tab_si:sensitivity_params} reports the moderator requirements and estimated costs to reach 80\% avoided hate engagement and 80\% coverage under all combinations of the two key operational parameters: reviewers per tweet (1 or 3) and review rate (100, 150, or 200 tweets per hour). Costs are reported for English alone, German alone, and all eight languages combined, matching the scope of the regimes considered in the main text. The primary finding is robust to these assumptions: achieving 80\% avoided hate engagement costs between 1.3\% and 7.9\% of annual revenue across all languages, between 0.2\% and 1.3\% for English alone, and between USD 3.4 and 20.7 million for German alone, so it remains an order of magnitude below the 10\% Online Safety Act ceiling and well below NetzDG's EUR 50 million cap under every parameter combination. Achieving 80\% coverage is substantially more expensive. Across all eight languages it costs 2.8\% to 16.6\% of revenue and exceeds the 10\% ceiling at the baseline review intensity, consistent with the main-text conclusion that comprehensive coverage is far less tractable than engagement-weighted targeting at current AI detection performance levels. For the two regulated languages taken on their own, however, even full coverage stays within the applicable ceilings (English 1.4\% to 8.2\% of revenue, German USD 2.3 to 13.6 million), because the languages that drive the all-language total are those where automated detection performs worst.

\input{tables/sensitivity_params_revised}

\subsubsection{Cost analysis}\label{si_cost_analysis}

\paragraph{Wage assumptions}
Annual cost per language was computed as $M_l \times w_l \times H$, where $M_l$ is the number of moderators allocated to language $l$, $w_l$ is the hourly wage for that language, and $H = 8 \times 365 = 2{,}920$ is the number of working hours per year per moderation position. The 365-day assumption reflects continuous platform operation: content is posted every day of the year, and maintaining a given number of active moderators per day across weekends and holidays requires staffing equivalent to 2,920 moderator-hours per position per year. $M_l$ therefore counts moderator positions rather than individual employees. Staffing a position around the clock requires more than one employee due to weekends, vacation, and sick leave, so the underlying employee headcount is somewhat higher than the position counts reported here. Because we cost positions rather than employees, the cost estimates are conservative lower bounds on the wage bill: realistic staffing with $\sim$15 days of annual leave per employee would imply $\sim$2{,}750 productive hours per position, i.e., approximately 6\% additional headcount, which would scale all reported costs upward by roughly the same factor.

Hourly wages were set using a purchasing-power-parity (PPP) adjustment. We took USD 20 as the hourly baseline for English-language moderation, consistent with the minimum wage floor Facebook set for U.S. content moderators in 2019 (ranging from USD 18/hr in mid-tier metros to USD 22/hr in San Francisco, New York, and Washington, DC; \citealp{frier2019facebook}) and at the lower end of subsequent salary surveys, which report U.S. averages between USD 19 and USD 28/hr for the role. Each language was then mapped to the most populous country in which it is the primary language (Arabic $\to$ Egypt, French $\to$ France, German $\to$ Germany, Indonesian $\to$ Indonesia, Portuguese $\to$ Brazil, Spanish $\to$ Mexico, Turkish $\to$ Turkey), and the USD 20 baseline was converted to the local PPP equivalent using 2022 World Bank International Comparison Program data \citep{WorldBank_WDI_PPP_2026}. Specifically, we computed the PPP-adjusted USD wage as USD~$20 \times (\text{PPP}_\ell / \text{FX}_\ell)$, where $\text{PPP}_\ell$ is the country's PPP conversion factor for private consumption (World Bank series PA.NUS.PRVT.PP) and $\text{FX}_\ell$ its market exchange rate (series PA.NUS.FCRF). The resulting per-language hourly wages are reported in table~\ref{tab_si:wage_assumptions}.

\input{tables/wage_assumptions}

Cost estimates are direct labor costs only and do not include overhead, management, training, or infrastructure. They should therefore be interpreted as conservative lower bounds on the true cost of operating a moderation system at the simulated scale.

\paragraph{Comparison to regulatory penalties}
Revenue was expressed as a share of Twitter's 2021 annual revenue, USD 5.08 billion \citep{Twitter_10K_2021}, the last full year reported before the platform went private in late 2022. The conclusions are unchanged under a lower 2022 estimate (USD 4.4 billion) or an aggregator figure (USD 5.22 billion).

Three regimes are relevant. At the time of our study, German content was governed by Germany's Network Enforcement Act (NetzDG) \citep{netzdg}, which set a fixed maximum fine of EUR 50 million (about USD 52.5 million in 2022) for systematic failure to remove illegal content, including hate speech, and has since been largely superseded by the DSA. The EU Digital Services Act (DSA) and UK Online Safety Act cap fines at 6\% and 10\% of global annual turnover respectively, the highest revenue-share penalties any content-moderation regulator levies. Our all-language estimate of 7.9\% of revenue to avoid 80\% of hate exposure falls below the 10\% OSA ceiling, though not significantly (one-sided $p=0.18$, below 10\% in 82\% of the 1{,}000 bootstrap resamples), and above the 6\% DSA ceiling.

This all-language figure overstates what any single jurisdiction requires, because it prices moderation of all eight languages worldwide and is driven by the languages where detection is hardest (Arabic and Turkish), which fall outside EU and UK jurisdiction. For the content each regime actually governs, the cost is far below the ceiling: moderating all English-language content to 80\% avoided engagement costs 1.3\% of revenue (versus 10\% under the OSA), German content about USD 21 million (95\% CI [8, 30] million, versus EUR 50 million under NetzDG), and French and German together 0.8\% (versus 6\% under the DSA). Because these price all worldwide content in each language, they are conservative upper bounds on single-jurisdiction compliance. We cover only two of the EU's twenty-four official languages, but the rest are generally well-resourced for detection, so full EU coverage would likely remain well below the 6\% ceiling.

These ceilings are statutory maxima, and actual penalties may be lower. Even so, enforcement is now active: in December 2025 the European Commission issued its first DSA fine, EUR 120 million against X \citep{CommissionFinesEU120}. The 10\% OSA base is itself contested, with Meta challenging Ofcom's revenue calculation in the UK High Court \citep{tobin2026meta}, which could narrow the base our comparison assumes.

%% file: tables/share_lang_twitterday.tex
\begin{table}[H]
    \centering
    \begin{tabular}{@{}lc@{}}
        \toprule
        \textbf{Language} & \textbf{Share (\%)} \\ \midrule
        \textbf{English}  & 27.6  \\
        Japanese  & 20.9  \\
        \textbf{Spanish}  &  6.7  \\
        \textbf{Arabic}  &  6.6  \\
        \textbf{Portuguese}  &  5.4  \\
        \textbf{Indonesian}  &  2.9  \\
        Korean  &  2.5  \\
        \textbf{Turkish}  &  2.2  \\
        Farsi  &  1.6  \\
        \textbf{French}  &  1.6  \\
        Thai  &  1.2  \\
        Tagalog  &  0.9  \\
        \textbf{German}  &  0.7  \\ \bottomrule
    \end{tabular}
    \caption{Share of all original tweets by language. Retained languages are bolded.}
    \label{tab:language_share_twitterday}
\end{table}

%% file: tables/kappa_3class.tex
\begin{table}[htbp]
\centering
\small
\begin{tabular}{lcc}
\hline
\textbf{Language / Country} & \textbf{N} & \textbf{Fleiss' $\kappa$ (3-class)} \\
\hline
\multicolumn{3}{l}{\textit{Languages}} \\
Arabic        & 45{,}000 & 0.426 \\
English       & 45{,}000 & 0.422 \\
French        & 45{,}000 & 0.510 \\
German        & 45{,}000 & 0.620 \\
Indonesian    & 45{,}000 & 0.427 \\
Portuguese    & 45{,}000 & 0.379 \\
Spanish       & 45{,}000 & 0.363 \\
Turkish       & 45{,}000 & 0.651 \\
\hline
\multicolumn{3}{l}{\textit{Countries (English-speaking)}} \\
India         & 45{,}000 & 0.627 \\
Kenya         & 45{,}000 & 0.505 \\
Nigeria       & 45{,}000 & 0.525 \\
United States & 45{,}000 & 0.641 \\
\hline
\end{tabular}
\caption{Inter-annotator agreement (Fleiss’ $\kappa$) for the primary three-way annotation.}
\label{tab:kappa_3class}
\end{table}

%% file: tables/kappa_violence.tex
\begin{table}[htbp]
\centering
\small
\begin{tabular}{lcc}
\hline
\textbf{Language / Country} & \textbf{N (hate items)} & \textbf{Cohen's $\kappa$} \\
\hline
\multicolumn{3}{l}{\textit{Languages}} \\
Arabic        & 162 & 0.610 \\
English       & 114 & 0.659 \\
French        & 314 & 0.788 \\
German        & 214 & 0.618 \\
Indonesian    & 84  & 0.794 \\
Portuguese    & 97  & 0.740 \\
Spanish       & 199 & 0.449 \\
Turkish       & 192 & 0.740 \\
\hline
\multicolumn{3}{l}{\textit{Countries (English-speaking)}} \\
India         & 294 & 0.608 \\
Kenya         & 117 & 0.487 \\
Nigeria       & 107 & 0.481 \\
United States & 137 & 0.386 \\
\hline
\end{tabular}
\caption{Inter-annotator agreement for violent vs.\ non-violent hate classification.}
\label{tab:kappa_violence}
\end{table}

%% file: tables/kappa_additional.tex
\begin{table}[htbp]
\centering
\small
\begin{tabular}{lccc}
\hline
\textbf{Language / Country} & \textbf{N} & \textbf{Scam $\kappa$} & \textbf{Adult $\kappa$} \\
\hline
\multicolumn{4}{l}{\textit{Languages}} \\
Arabic        & 25{,}000 & 0.388 & 0.817 \\
English       & 25{,}000 & 0.385 & 0.560 \\
French        & 25{,}000 & 0.244 & 0.641 \\
German        & 25{,}000 & 0.169 & 0.637 \\
Indonesian    & 25{,}000 & 0.129 & 0.741 \\
Portuguese    & 25{,}000 & 0.304 & 0.625 \\
Spanish       & 25{,}000 & 0.158 & 0.712 \\
Turkish       & 25{,}000 & 0.223 & 0.917 \\
\hline
\multicolumn{4}{l}{\textit{Countries (English-speaking)}} \\
India         & 25{,}000 & 0.619 & 0.745 \\
Kenya         & 25{,}000 & 0.295 & 0.370 \\
Nigeria       & 25{,}000 & 0.150 & 0.341 \\
United States & 25{,}000 & 0.247 & 0.657 \\
\hline
\end{tabular}
\caption{Inter-annotator agreement (Fleiss’ $\kappa$) for additional content categories.}
\label{tab:kappa_additional}
\end{table}

%% file: figures/prevalence_composition_stats.tex

\begin{figure}[H]
  \centering
    \includegraphics[width=\textwidth]{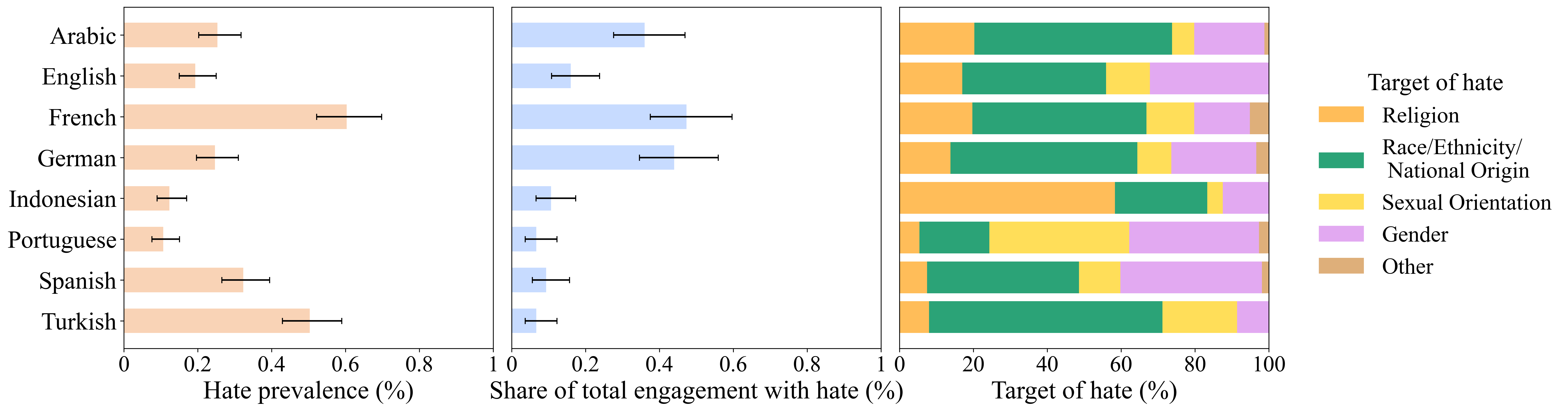}
  \caption{Prevalence, share of total engagement and target composition of hate speech across languages. Error bars indicate 95\% bootstrapped confidence intervals.}
  \label{fig:prevalence_composition_lang}
\end{figure}

\begin{figure}[H]
  \centering
    \includegraphics[width=\textwidth]{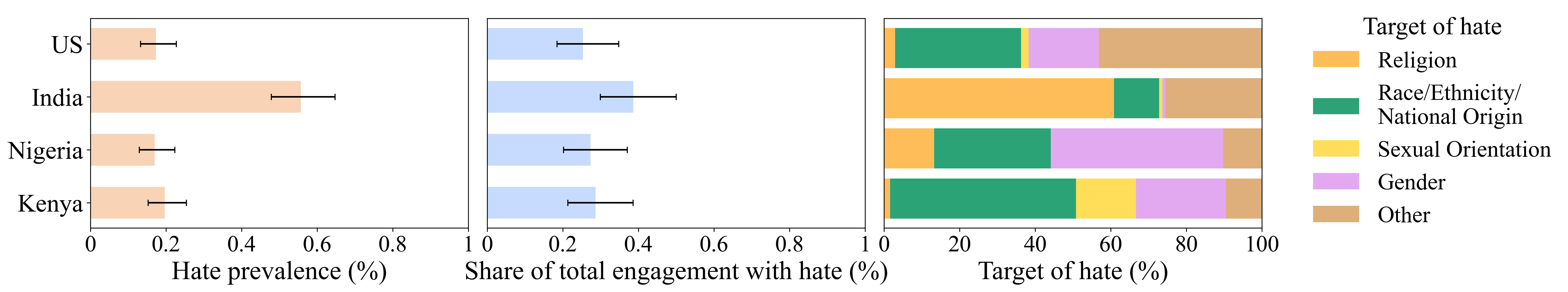}
  \caption{Prevalence, share of total engagement and target composition of hate speech across countries. Error bars indicate 95\% bootstrapped confidence intervals.}
  \label{fig:prevalence_composition_country}
\end{figure}

%% file: tables/per_language_exposure.tex
\begin{table}[htbp]
\centering
\small
\begin{tabular}{lccc}
\hline
\textbf{Language} & \textbf{Prevalence (\%)} & \textbf{Hateful tweets/day} & \textbf{$\approx$ 1 per} \\
\hline
French     & 0.60 & 1.21 & 0.8 days \\
Turkish    & 0.50 & 1.01 & 1.0 days \\
Spanish    & 0.32 & 0.65 & 1.5 days \\
Arabic     & 0.25 & 0.51 & 2.0 days \\
German     & 0.25 & 0.49 & 2.0 days \\
English    & 0.19 & 0.39 & 2.6 days \\
Indonesian & 0.12 & 0.25 & 4.1 days \\
Portuguese & 0.11 & 0.21 & 4.7 days \\
\hline
\end{tabular}
\caption{Expected daily exposure to hate for a monolingual user, by language. $\hat{p}_\ell$ is the Twitter-policy hate prevalence in the random sample. $E_\ell = \hat{p}_\ell \times F$ with $F=200$ tweets viewed per day.}
\label{tab_si:per_language_exposure}
\end{table}

%% file: tables/count_removal_hate.tex
\begin{table}[H]
\centering
\scriptsize
\begin{tabular}{lccccccccccc}
\hline
\textbf{Language/Country} & \textbf{N} &
\multicolumn{3}{c}{\textbf{Total Hate}} &
\multicolumn{3}{c}{\textbf{Non-Violent Hate}} &
\multicolumn{3}{c}{\textbf{Violent Hate}} \\
\cline{3-11}
 &  &
\textbf{Count} & \textbf{Removed} & \textbf{Rate (\%)} &
\textbf{Count} & \textbf{Removed} & \textbf{Rate (\%)} &
\textbf{Count} & \textbf{Removed} & \textbf{Rate (\%)} \\
\hline
\textbf{Total (Languages)} & 240000 &
706 & 144 & 20.40 &
646 & 132 & 20.43 &
60 & 12 & 20.00 \\
\hline
Arabic        & 30000 & 76 & 17 & 22.37 & 70 & 17 & 24.29 & 6 & 0 & 0.00 \\
English       & 30000 & 58 & 15 & 25.86 & 57 & 15 & 26.32 & 1 & 0 & 0.00 \\
French        & 30000 & 181 & 36 & 19.89 & 167 & 33 & 19.76 & 14 & 3 & 21.43 \\
German        & 30000 & 74 & 12 & 16.22 & 68 & 9 & 13.24 & 6 & 3 & 50.00 \\
Indonesian    & 30000 & 37 & 9 & 24.32 & 35 & 8 & 22.86 & 2 & 1 & 50.00 \\
Portuguese    & 30000 & 32 & 3 & 9.38 & 30 & 3 & 10.00 & 2 & 0 & 0.00 \\
Spanish       & 30000 & 97 & 17 & 17.53 & 93 & 16 & 17.20 & 4 & 1 & 25.00 \\
Turkish       & 30000 & 151 & 35 & 23.18 & 126 & 31 & 24.60 & 25 & 4 & 16.00 \\
\hline
\textbf{Total (Countries)} & 120000 &
329 & 50 & 15.20 &
306 & 49 & 16.01 &
23 & 1 & 4.35 \\
\hline
India         & 30000 & 167 & 32 & 19.16 & 148 & 32 & 21.62 & 19 & 0 & 0.00 \\
Kenya         & 30000 & 59 & 6 & 10.17 & 58 & 6 & 10.34 & 1 & 0 & 0.00 \\
Nigeria       & 30000 & 51 & 4 & 7.84 & 50 & 3 & 6.00 & 1 & 1 & 100.00 \\
United States & 30000 & 52 & 8 & 15.38 & 50 & 8 & 16.00 & 2 & 0 & 0.00 \\
\hline
\end{tabular}
\caption{Counts, removals, and removal rates (percent) for hate-related variables across languages and countries, including aggregate totals.}
\label{tab_si:count_removal_hate}
\end{table}

%% file: tables/count_removal_problematic.tex
\begin{table}[H]
\centering
\scriptsize
\begin{tabular}{lcccccccc}
\hline
\textbf{Language/Country} & \textbf{N} &
\multicolumn{3}{c}{\textbf{Scam}} &
\multicolumn{3}{c}{\textbf{Adult Content}} \\
\cline{3-8}
 &  &
\textbf{Count} & \textbf{Removed} & \textbf{Rate (\%)} &
\textbf{Count} & \textbf{Removed} & \textbf{Rate (\%)} \\
\hline
\textbf{Total (Languages)} & 240000 &
147 & 83 & 56.46 &
1269 & 635 & 50.04 \\
\hline
Arabic        & 30000 & 18 & 8 & 44.44 & 185 & 109 & 58.92 \\
English       & 30000 & 39 & 20 & 51.28 & 96 & 29 & 30.21 \\
French        & 30000 & 4 & 4 & 100.00 & 37 & 9 & 24.32 \\
German        & 30000 & 6 & 2 & 33.33 & 73 & 22 & 30.14 \\
Indonesian    & 30000 & 10 & 2 & 20.00 & 57 & 15 & 26.32 \\
Portuguese    & 30000 & 42 & 36 & 85.71 & 128 & 26 & 20.31 \\
Spanish       & 30000 & 8 & 0 & 0.00 & 148 & 30 & 20.27 \\
Turkish       & 30000 & 20 & 11 & 55.00 & 545 & 395 & 72.48 \\
\hline
\textbf{Total (Countries)} & 120000 &
192 & 73 & 38.02 &
231 & 34 & 14.72 \\
\hline
India         & 30000 & 66 & 15 & 22.73 & 79 & 10 & 12.66 \\
Kenya         & 30000 & 63 & 19 & 30.16 & 20 & 2 & 10.00 \\
Nigeria       & 30000 & 33 & 12 & 36.36 & 36 & 3 & 8.33 \\
United States & 30000 & 30 & 27 & 90.00 & 96 & 19 & 19.79 \\
\hline
\end{tabular}
\caption{Counts, removals, and removal rates (percent) for non-hate problematic content categories (scam and adult material) across languages and countries, including aggregate totals.}
\label{tab_si:count_removal_problematic}
\end{table}

%% file: tables/count_removal_misc.tex
\begin{table}[H]
\centering
\scriptsize
\begin{tabular}{lccccccccccc}
\hline
\textbf{Language/Country} & \textbf{N} &
\multicolumn{3}{c}{\textbf{Sensitive Media}} &
\multicolumn{3}{c}{\textbf{Is Reply}} &
\multicolumn{3}{c}{\textbf{Author Verified}} \\
\cline{3-11}
 &  &
\textbf{Count} & \textbf{Removed} & \textbf{Rate (\%)} &
\textbf{Count} & \textbf{Removed} & \textbf{Rate (\%)} &
\textbf{Count} & \textbf{Removed} & \textbf{Rate (\%)} \\
\hline
\textbf{Total (Languages)} & 240000 &
1958 & 510 & 26.05 &
137160 & 18217 & 13.28 &
3273 & 56 & 1.71 \\
\hline
Arabic        & 30000 & 201 & 74 & 36.82 & 14447 & 2158 & 14.94 & 196 & 5 & 2.55 \\
English       & 30000 & 372 & 101 & 27.15 & 16596 & 2447 & 14.74 & 527 & 9 & 1.71 \\
French        & 30000 & 186 & 45 & 24.19 & 19927 & 2314 & 11.61 & 380 & 10 & 2.63 \\
German        & 30000 & 217 & 64 & 29.49 & 21401 & 2772 & 12.95 & 471 & 9 & 1.91 \\
Indonesian    & 30000 & 306 & 69 & 22.55 & 17134 & 2136 & 12.47 & 874 & 2 & 0.23 \\
Portuguese    & 30000 & 177 & 34 & 19.21 & 14963 & 1858 & 12.42 & 163 & 6 & 3.68 \\
Spanish       & 30000 & 298 & 58 & 19.46 & 16665 & 1206 & 7.24 & 460 & 13 & 2.83 \\
Turkish       & 30000 & 201 & 65 & 32.34 & 16027 & 3326 & 20.75 & 202 & 2 & 0.99 \\
\hline
\textbf{Total (Countries)} & 120000 &
1235 & 225 & 18.22 &
76058 & 6768 & 8.90 &
3976 & 68 & 1.71 \\
\hline
India         & 30000 & 366 & 51 & 13.93 & 16949 & 1647 & 9.72 & 1560 & 30 & 1.92 \\
Kenya         & 30000 & 119 & 14 & 11.76 & 20486 & 1294 & 6.32 & 817 & 5 & 0.61 \\
Nigeria       & 30000 & 244 & 36 & 14.75 & 21495 & 1876 & 8.73 & 284 & 3 & 1.06 \\
United States & 30000 & 506 & 124 & 24.51 & 17128 & 1951 & 11.39 & 1315 & 30 & 2.28 \\
\hline
\end{tabular}
\caption{Counts, removals, and removal rates (percent) for sensitive media, replies, and verified authors across languages and countries, including aggregate totals.}
\label{tab_si:count_removal_misc}
\end{table}

%% file: tables/tweet_level_metrics.tex
\begin{table}[H]
\centering
\scriptsize
\begin{tabular}{lccc}
\hline
\textbf{Language/Country} &
\textbf{Retweets} &
\textbf{Sensitive (\%)} &
\textbf{Reply (\%)} \\
\hline
\textbf{Total (Languages)} 
& 0.47 (4.27) & 0.82 & 57.15 \\
\hline
Arabic        & 0.11 (1.57)  & 0.67 & 48.16 \\
German        & 0.07 (0.73)  & 0.72 & 71.34 \\
English       & 0.16 (1.93)  & 1.24 & 55.32 \\
Spanish       & 0.12 (1.42)  & 0.99 & 55.55 \\
French        & 0.12 (3.81)  & 0.62 & 66.42 \\
Indonesian    & 0.06 (2.17)  & 1.02 & 57.11 \\
Portuguese    & 0.07 (1.14)  & 0.59 & 49.88 \\
Turkish       & 3.00 (10.47) & 0.67 & 53.42 \\
\hline
\textbf{Total (Countries)} 
& 0.31 (7.60) & 1.03 & 63.38 \\
\hline
India         & 0.67 (14.78) & 1.22 & 56.50 \\
Kenya         & 0.27 (2.02)  & 0.40 & 68.29 \\
Nigeria       & 0.18 (2.29)  & 0.81 & 71.65 \\
United States & 0.13 (1.80)  & 1.69 & 57.09 \\
\hline
\end{tabular}
\caption{Descriptive statistics for tweet-level covariates used in removal regressions. Retweets are reported as mean values with standard deviations in parentheses. Sensitive and Reply indicate the percentage of tweets flagged as sensitive or posted as replies.}
\label{tab_si:tweet_covariates}
\end{table}

%% file: tables/user_level_metrics.tex
\begin{table}[H]
\centering
\scriptsize
\begin{tabular}{lcccc}
\hline
\textbf{Language/Country} &
\textbf{Verified (\%)} &
\textbf{Followers (median)} &
\textbf{Following (median)} &
\textbf{Tweet Count (median)} \\
\hline
\textbf{Total (Languages)}
& 1.36 & 214 & 270 & 4{,}979 \\
\hline
Arabic        & 0.65 & 72  & 130 & 2{,}108 \\
German        & 1.57 & 233 & 282 & 6{,}289 \\
English       & 1.76 & 241 & 321 & 5{,}289 \\
Spanish       & 1.53 & 265 & 332 & 5{,}987 \\
French        & 1.27 & 244 & 315 & 7{,}031 \\
Indonesian    & 2.91 & 249 & 255 & 7{,}192 \\
Portuguese    & 0.54 & 237 & 305 & 5{,}370 \\
Turkish       & 0.67 & 160 & 207 & 2{,}019 \\
\hline
\textbf{Total (Countries)}
& 3.31 & 516 & 544 & 5{,}806 \\
\hline
India         & 5.20 & 281 & 292 & 5{,}811 \\
Kenya         & 2.72 & 828 & 728 & 4{,}717 \\
Nigeria       & 0.95 & 557 & 682 & 4{,}317 \\
United States & 4.38 & 480 & 560 & 8{,}880 \\
\hline
\end{tabular}
\caption{Descriptive statistics for user-level covariates used in regression models. Continuous variables are reported as medians. Verified indicates the percentage of verified accounts. Tweet Count is the total number of statuses posted on the user's account at the time of collection (lifetime \texttt{statuses\_count}), not the count of tweets posted within the 24-hour snapshot.}
\label{tab_si:user_covariates}
\end{table}

%% file: tables/user_suspension_rates.tex
\begin{table}[H]
\centering
\scriptsize
\begin{tabular}{lccccccccc}
\hline
\textbf{Language/Country} & \textbf{N} &
\multicolumn{2}{c}{\textbf{Overall Suspension}} &
\multicolumn{3}{c}{\textbf{Hateful Users}} &
\multicolumn{3}{c}{\textbf{Non-Hateful Users}} \\
\cline{3-4} \cline{5-7} \cline{8-10}
 &  &
\textbf{Susp.} & \textbf{Rate (\%)} &
\textbf{Count} & \textbf{Susp.} & \textbf{Rate (\%)} &
\textbf{Count} & \textbf{Susp.} & \textbf{Rate (\%)} \\
\hline
\textbf{Total (Languages)} & 215{,}594 &
17{,}638 & 8.18 &
702 & 59 & 8.40 &
214{,}892 & 17{,}579 & 8.18 \\
\hline
Arabic        & 26{,}035 & 7{,}089 & 27.23 & 75  & 9  & 12.00 & 25{,}960 & 7{,}080 & 27.27 \\
German        & 24{,}303 & 1{,}209 & 4.97  & 73  & 6  & 8.22  & 24{,}230 & 1{,}203 & 4.96  \\
English       & 29{,}251 & 3{,}096 & 10.58 & 58  & 7  & 12.07 & 29{,}193 & 3{,}089 & 10.58 \\
Spanish       & 29{,}173 & 622     & 2.13  & 97  & 6  & 6.19  & 29{,}076 & 616     & 2.12  \\
French        & 27{,}073 & 1{,}001 & 3.70  & 180 & 17 & 9.44  & 26{,}893 & 984     & 3.66  \\
Indonesian    & 25{,}654 & 1{,}368 & 5.33  & 36  & 6  & 16.67 & 25{,}618 & 1{,}362 & 5.32  \\
Portuguese    & 29{,}069 & 799     & 2.75  & 32  & 1  & 3.12  & 29{,}037 & 798     & 2.75  \\
Turkish       & 25{,}036 & 2{,}454 & 9.80  & 151 & 7  & 4.64  & 24{,}885 & 2{,}447 & 9.83  \\
\hline
\textbf{Total (Countries)} & 84{,}206 &
2{,}751 & 3.27 &
320 & 28 & 8.75 &
83{,}886 & 2{,}723 & 3.25 \\
\hline
India         & 22{,}097 & 687  & 3.11 & 161 & 19 & 11.80 & 21{,}936 & 668 & 3.05 \\
Kenya         & 13{,}760 & 295  & 2.14 & 57  & 3  & 5.26  & 13{,}703 & 292 & 2.13 \\
Nigeria       & 20{,}493 & 699  & 3.41 & 50  & 1  & 2.00  & 20{,}443 & 698 & 3.41 \\
United States & 27{,}856 & 1{,}070 & 3.84 & 52  & 5  & 9.62  & 27{,}804 & 1{,}065 & 3.83 \\
\hline
\end{tabular}
\caption{User-level suspension rates across languages and countries. A user is classified as suspended if at least one tweet associated with the account was returned with reason \texttt{suspended} in the Compliance API. Hateful users are defined as users who posted at least one tweet labeled as hate under Twitter’s hateful conduct policy.}
\label{tab_si:user_suspension_rates}
\end{table}

%% file: tables/count_removal_hate_by_reason.tex
\begin{table}[H]
\centering
\scriptsize
\begin{tabular}{lcccccc}
\hline
\textbf{Language/Country} & \textbf{N} & \textbf{Hate count} &
\textbf{Total removed (\%)} & \textbf{Suspended (\%)} &
\textbf{Deactivated (\%)} & \textbf{Deleted (\%)} \\
\hline
\textbf{Total (Languages)} & 240{,}000 & 706 & 20.40 &  8.64 & 0.71 & 11.76 \\
\hline
Arabic        & 30{,}000 &  76 & 22.37 & 13.16 & 0.00 &  9.21 \\
English       & 30{,}000 &  58 & 25.86 & 12.07 & 0.00 & 13.79 \\
French        & 30{,}000 & 181 & 19.89 &  9.39 & 1.66 & 10.50 \\
German        & 30{,}000 &  74 & 16.22 &  9.46 & 0.00 &  6.76 \\
Indonesian    & 30{,}000 &  37 & 24.32 & 16.22 & 0.00 &  8.11 \\
Portuguese    & 30{,}000 &  32 &  9.38 &  3.12 & 0.00 &  6.25 \\
Spanish       & 30{,}000 &  97 & 17.53 &  6.19 & 0.00 & 11.34 \\
Turkish       & 30{,}000 & 151 & 23.18 &  4.64 & 1.32 & 18.54 \\
\hline
\textbf{Total (Countries)} & 120{,}000 & 329 & 15.20 &  8.81 & 0.30 &  6.38 \\
\hline
India         & 30{,}000 & 167 & 19.16 & 11.98 & 0.00 &  7.19 \\
Kenya         & 30{,}000 &  59 & 10.17 &  5.08 & 0.00 &  5.08 \\
Nigeria       & 30{,}000 &  51 &  7.84 &  1.96 & 0.00 &  5.88 \\
United States & 30{,}000 &  52 & 15.38 &  9.62 & 1.92 &  5.77 \\
\hline
\end{tabular}
\caption{Decomposition of hate-tweet removal rates by removal reason, across languages and countries (random samples, $N=30{,}000$ per language/country). ``Suspended'' refers to tweets removed because the posting account was suspended (unambiguously platform-initiated); ``Deleted'' to removal of an individual tweet without account-level action (the Compliance API does not distinguish platform- from user-initiated tweet deletion). ``Deactivated'' refers to user-initiated account deactivation, which is \emph{not} counted as removed under our definition and is shown here only for completeness; protected (private) accounts are likewise excluded. The ``Total removed'' column equals the sum of the Suspended and Deleted columns (within rounding); suspension constitutes a lower bound on platform-initiated enforcement.}
\label{tab_si:count_removal_hate_by_reason}
\end{table}

%% file: tables/tweet_level_removal_reg_lang_claude.tex
\begin{table}[H]
\centering
\small
\begin{tabular}{lcccc}
\hline
 & \multicolumn{4}{c}{Tweet Removed (Language Sample)} \\
 & (1) Hate & (2) Violent vs.\ Non-Violent Hate & (3) + User Controls & (4) + Policy Categories \\
\hline
\textbf{Hate Indicators (vs.\ Neutral)} & & & & \\

$\quad$ Hate
& 0.03$^{\dagger}$
&  &  &  \\
& (0.01)
&  &  &  \\

$\quad$ Violent Hate
& 
& 0.01
& 0.00
& -0.06 \\
& 
& (0.05)
& (0.05)
& (0.08) \\

$\quad$ Non-Violent Hate
& 
& 0.03$^{\dagger}$
& 0.02
& 0.01 \\
& 
& (0.01)
& (0.01)
& (0.02) \\

\textbf{Tweet-Level Controls} & & & & \\

$\quad$ Retweet Count (log)
& 0.04$^{***}$
& 0.04$^{***}$
& 0.05$^{***}$
& 0.04$^{***}$ \\
& (0.00)
& (0.00)
& (0.00)
& (0.00) \\

$\quad$ Possibly Sensitive
& 0.05$^{***}$
& 0.05$^{***}$
& 0.08$^{***}$
& 0.04$^{**}$ \\
& (0.01)
& (0.01)
& (0.01)
& (0.01) \\

$\quad$ Is Reply
& -0.10$^{***}$
& -0.10$^{***}$
& -0.07$^{***}$
& -0.07$^{***}$ \\
& (0.00)
& (0.00)
& (0.00)
& (0.00) \\

\textbf{User-Level Controls} & & & & \\

$\quad$ Verified Author
&  &  & -0.02$^{*}$ & -0.01 \\
&  &  & (0.01) & (0.01) \\

$\quad$ Followers (log)
&  &  & -0.01$^{***}$ & -0.01$^{***}$ \\
&  &  & (0.00) & (0.00) \\

$\quad$ Following (log)
&  &  & -0.02$^{***}$ & -0.02$^{***}$ \\
&  &  & (0.00) & (0.00) \\

$\quad$ User Tweet Count (log)
&  &  & 0.01$^{***}$ & 0.01$^{***}$ \\
&  &  & (0.00) & (0.00) \\

$\quad$ Account Age (log)
&  &  & -0.04$^{***}$ & -0.04$^{***}$ \\
&  &  & (0.00) & (0.00) \\

\textbf{Additional Policy Categories} & & & & \\

$\quad$ Scam
&  &  &  & 0.29$^{***}$ \\
&  &  &  & (0.03) \\

$\quad$ Adult Content
&  &  &  & 0.17$^{***}$ \\
&  &  &  & (0.01) \\

\hline
Intercept (Neutral Baseline)
& 0.45$^{***}$
& 0.45$^{***}$
& 0.74$^{***}$
& 0.73$^{***}$ \\
& (0.00)
& (0.00)
& (0.00)
& (0.01) \\

\hline
Fixed Effects
& Yes & Yes & Yes & Yes \\

Observations
& 240,000
& 240,000
& 240,000
& 80,000 \\

$R^2$
& 0.08
& 0.08
& 0.17
& 0.18 \\

\hline
\multicolumn{5}{l}{Standard errors in parentheses (OLS).} \\
\multicolumn{5}{l}{$^{***}p<0.001$, $^{**}p<0.01$, $^{*}p<0.05$, $^{\dagger}p<0.10$.} \\
\end{tabular}
\caption{Linear probability models predicting tweet removal five months after posting (language sample). All specifications include language fixed effects (English is the omitted reference). Column (1) includes a single Twitter-defined hate indicator. Column (2) separates violent and non-violent Twitter hate. Column (3) additionally includes user-level controls (verified, log followers, log following, log lifetime tweet count, log account age) on the full sample. Column (4) further adds other policy-relevant content categories (scam, adult content); its smaller sample reflects missing values for those categories.}
\label{tab_si:tweet_removal_language_reg}
\end{table}

%% file: tables/tweet_level_removal_reg_country_claude.tex
\begin{table}[H]
\centering
\small
\begin{tabular}{lcccc}
\hline
 & \multicolumn{4}{c}{Tweet Removed (Country Sample)} \\
 & (1) Hate & (2) Violent vs.\ Non-Violent Hate & (3) + User Controls & (4) + Policy Categories \\
\hline
\textbf{Hate Indicators (vs.\ Neutral)} & & & & \\

$\quad$ Hate
& 0.06$^{***}$
&  &  &  \\
& (0.02)
&  &  &  \\

$\quad$ Violent Hate
& 
& -0.06
& -0.06
& -0.11 \\
& 
& (0.06)
& (0.06)
& (0.10) \\

$\quad$ Non-Violent Hate
& 
& 0.07$^{***}$
& 0.06$^{***}$
& 0.05$^{\dagger}$ \\
& 
& (0.02)
& (0.02)
& (0.03) \\

\textbf{Tweet-Level Controls} & & & & \\

$\quad$ Retweet Count (log)
& -0.00
& -0.00
& 0.01$^{***}$
& 0.01 \\
& (0.00)
& (0.00)
& (0.00)
& (0.00) \\

$\quad$ Possibly Sensitive
& 0.08$^{***}$
& 0.08$^{***}$
& 0.07$^{***}$
& 0.08$^{***}$ \\
& (0.01)
& (0.01)
& (0.01)
& (0.02) \\

$\quad$ Is Reply
& -0.02$^{***}$
& -0.02$^{***}$
& -0.02$^{***}$
& -0.02$^{***}$ \\
& (0.00)
& (0.00)
& (0.00)
& (0.00) \\

\textbf{User-Level Controls} & & & & \\

$\quad$ Verified Author
&  &  & -0.04$^{***}$ & -0.05$^{***}$ \\
&  &  & (0.01) & (0.01) \\

$\quad$ Followers (log)
&  &  & -0.00 & 0.00 \\
&  &  & (0.00) & (0.00) \\

$\quad$ Following (log)
&  &  & -0.01$^{***}$ & -0.01$^{***}$ \\
&  &  & (0.00) & (0.00) \\

$\quad$ User Tweet Count (log)
&  &  & -0.01$^{***}$ & -0.01$^{***}$ \\
&  &  & (0.00) & (0.00) \\

$\quad$ Account Age (log)
&  &  & -0.02$^{***}$ & -0.02$^{***}$ \\
&  &  & (0.00) & (0.00) \\

\textbf{Additional Policy Categories} & & & & \\

$\quad$ Scam
&  &  &  & 0.26$^{***}$ \\
&  &  &  & (0.02) \\

$\quad$ Adult Content
&  &  &  & 0.01 \\
&  &  &  & (0.02) \\

\hline
Intercept (Neutral Baseline)
& 0.12$^{***}$
& 0.12$^{***}$
& 0.35$^{***}$
& 0.36$^{***}$ \\
& (0.00)
& (0.00)
& (0.01)
& (0.01) \\

\hline
Fixed Effects
& Yes & Yes & Yes & Yes \\

Observations
& 120,000
& 120,000
& 120,000
& 40,000 \\

$R^2$
& 0.00
& 0.00
& 0.03
& 0.03 \\

\hline
\multicolumn{5}{l}{Standard errors in parentheses (OLS).} \\
\multicolumn{5}{l}{$^{***}p<0.001$, $^{**}p<0.01$, $^{*}p<0.05$, $^{\dagger}p<0.10$.} \\
\end{tabular}
\caption{Linear probability models predicting tweet removal in the country-restricted sample (US, India, Nigeria, Kenya). All specifications include country fixed effects (US is the omitted reference). Column (1) includes a general Twitter-defined hate indicator. Column (2) separates violent and non-violent Twitter hate. Column (3) additionally includes user-level controls (verified, log followers, log following, log lifetime tweet count, log account age) on the full sample. Column (4) further adds other policy-relevant content categories (scam, adult content); its smaller sample reflects missing values for those categories.}
\label{tab_si:tweet_removal_country_reg}
\end{table}

%% file: tables/tweet_level_removal_reg_interaction_lang_claude.tex
\begin{table}[H]
\centering
\small
\begin{tabular}{lcc}
\hline
 & \multicolumn{2}{c}{Tweet Removed (Language Sample)} \\
 & (1) Full controls & (2) + Policy categories \\
\hline
\textbf{Main Effects} & & \\

Hate (English baseline) & 0.06 & 0.04 \\
& (0.06) & (0.10) \\

\textbf{Language Main Effects (vs English)} & & \\
Arabic & 0.12$^{***}$ & 0.12$^{***}$ \\
German & -0.03$^{***}$ & -0.03$^{***}$ \\
Spanish & -0.07$^{***}$ & -0.07$^{***}$ \\
French & -0.04$^{***}$ & -0.03$^{***}$ \\
Indonesian & -0.05$^{***}$ & -0.04$^{***}$ \\
Portuguese & -0.05$^{***}$ & -0.04$^{***}$ \\
Turkish & 0.06$^{***}$ & 0.06$^{***}$ \\

\textbf{Additional Policy Categories} & & \\
Scam &  & 0.29$^{***}$ \\
Adult Content &  & 0.17$^{***}$ \\

\textbf{Interaction Terms (Hate $\times$ Language, ref = English)} & & \\
$\quad$Arabic & -0.14$^{\dagger}$ & -0.15 \\
$\quad$German & -0.05 & -0.05 \\
$\quad$Spanish & -0.02 & 0.02 \\
$\quad$French & -0.02 & 0.02 \\
$\quad$Indonesian & 0.07 & 0.19 \\
$\quad$Portuguese & -0.09 & -0.15 \\
$\quad$Turkish & -0.08 & -0.10 \\
\hline
Joint Wald (all interactions $=0$) & $F_{7}=1.52$, $p=0.155$ & $F_{7}=2.16$, $p=0.034$ \\
Observations & 240,000 & 80,000 \\
$R^2$ & 0.17 & 0.18 \\
\hline
\multicolumn{3}{l}{\footnotesize $^{***}p<0.001$, $^{**}p<0.01$, $^{*}p<0.05$, $^{\dagger}p<0.10$.} \\
\end{tabular}
\caption{Linear probability models predicting tweet removal with an interaction between Twitter-defined hate and language (Language Sample, new removal definition, heteroskedasticity-robust HC1 standard errors in parentheses). English is the reference. Both columns use the full column-(4) control set of table~\ref{tab_si:tweet_removal_language_reg}, and column (2) additionally includes scam and adult-content indicators (policy-annotated subsample). Control coefficients are suppressed for compactness.}
\label{tab_si:language_interaction_removal}
\end{table}

%% file: tables/tweet_level_removal_reg_interaction_country_claude.tex
\begin{table}[H]
\centering
\small
\begin{tabular}{lcc}
\hline
 & \multicolumn{2}{c}{Tweet Removed (Country Sample)} \\
 & (1) Full controls & (2) + Policy categories \\
\hline
\textbf{Main Effects} & & \\

Hate (United States baseline) & 0.03 & 0.01 \\
& (0.05) & (0.08) \\

\textbf{Country Main Effects (vs United States)} & & \\
India & -0.03$^{***}$ & -0.03$^{***}$ \\
Nigeria & -0.03$^{***}$ & -0.03$^{***}$ \\
Kenya & -0.05$^{***}$ & -0.06$^{***}$ \\

\textbf{Additional Policy Categories} & & \\
Scam &  & 0.26$^{***}$ \\
Adult Content &  & 0.01 \\

\textbf{Interaction Terms (Hate $\times$ Country, ref = United States)} & & \\
$\quad$India & 0.07 & 0.10 \\
$\quad$Nigeria & -0.05 & -0.07 \\
$\quad$Kenya & -0.00 & 0.01 \\
\hline
Joint Wald (all interactions $=0$) & $F_{3}=2.13$, $p=0.095$ & $F_{3}=1.50$, $p=0.214$ \\
Observations & 120,000 & 40,000 \\
$R^2$ & 0.03 & 0.03 \\
\hline
\multicolumn{3}{l}{\footnotesize $^{***}p<0.001$, $^{**}p<0.01$, $^{*}p<0.05$, $^{\dagger}p<0.10$.} \\
\end{tabular}
\caption{Linear probability models predicting tweet removal with an interaction between Twitter-defined hate and country (Country Sample, new removal definition, heteroskedasticity-robust HC1 standard errors in parentheses). United States is the reference. Both columns use the full column-(4) control set of table~\ref{tab_si:tweet_removal_country_reg}, and column (2) additionally includes scam and adult-content indicators (policy-annotated subsample). Control coefficients are suppressed for compactness.}
\label{tab_si:country_interaction_removal}
\end{table}

%% file: tables/removal_equivalence_power.tex
\begin{table}[H]
\centering
\small
\begin{tabular}{lrrccc}
\hline
Estimate & $n_{\text{hate}}$ & $\hat{\beta}$ (pp) & 95\% CI (pp) & TOST bound & MDE \\
\hline
Pooled hate, language & 706 & $+2.6$ & $[-0.4,\ +5.6]$ & $\pm5.2$ & $4.4$ \\
Pooled hate, country & 329 & $+5.5$ & $[+1.6,\ +9.5]$ & --- & $5.6$ \\
Non-violent hate, language & 646 & $+2.8$ & $[-0.4,\ +6.0]$ & $\pm5.5$ & $4.5$ \\
Violent hate, language & 60 & $+0.4$ & $[-10.3,\ +11.1]$ & $\pm9.3$ & $15.3$ \\
Hate $\times$ language (joint) & 706 & \multicolumn{2}{c}{$F_{7}=1.52,\ p=0.15$} & --- & $18$--$26$ \\
Hate $\times$ country (joint) & 329 & \multicolumn{2}{c}{$F_{3}=2.13,\ p=0.09$} & --- & $16$--$18$ \\
\hline
\end{tabular}
\caption{Verified confidence intervals, equivalence (TOST) bounds, and minimum detectable effects for the tweet-removal nulls. All quantities are in percentage points, computed with HC1 robust standard errors. The TOST bound is the premium magnitude rejected by two one-sided tests at $\alpha=0.05$, and is omitted where the equivalence test does not apply (the country row and the joint interaction rows). MDE is the minimum effect detectable at 80\% power, and for the interaction rows it is the range of per-context contrast MDEs. The hate, violent hate, and non-violent hate rows are the columns-(1) and (2) specifications on the full language and country samples, without policy-category controls, and therefore differ from the columns-(3)/(4) estimates in table~\ref{tab_si:tweet_removal_language_reg}, which add scam and adult-content controls on the smaller policy-annotated subsample.}
\label{tab_si:removal_equivalence_power}
\end{table}

%% file: tables/user_level_suspension_reg_language.tex
\begin{table}[H]
\centering
\small
\begin{tabular}{lc}
\hline
 & User Suspended \\
\hline
User Posted Hate
& $-0.0054$ \\
& $(0.0091)$ \\

Tweet Count (log)
& $0.2235^{***}$ \\
& $(0.0034)$ \\

Followers (log)
& $-0.00442^{***}$ \\
& $(0.00034)$ \\

Following (log)
& $-0.02547^{***}$ \\
& $(0.00042)$ \\

Account Age (log)
& $-0.0302^{***}$ \\
& $(0.0003)$ \\

Verified
& $0.0099^{\dagger}$ \\
& $(0.0056)$ \\

Intercept
& $0.3935^{***}$ \\
& $(0.0034)$ \\

\hline
Language Fixed Effects & Yes \\
Observations & 214,624 \\
$R^2$ & 0.223 \\
\hline
\multicolumn{2}{l}{Standard errors in parentheses (OLS).} \\
\end{tabular}
\caption{User-level linear probability model predicting suspension (language sample) with language fixed effects.}
\label{tab_si:user_suspension_language_reg}

\end{table}

%% file: tables/user_level_suspension_reg_country.tex
\begin{table}[H]
\centering
\small
\begin{tabular}{lc}
\hline
 & User Suspended \\
\hline
User Posted Hate
& $0.0506^{***}$ \\
& $(0.0098)$ \\

Tweet Count (log)
& $0.0442^{***}$ \\
& $(0.0021)$ \\

Followers (log)
& $-0.00291^{***}$ \\
& $(0.00039)$ \\

Following (log)
& $-0.00128^{**}$ \\
& $(0.00049)$ \\

Account Age (log)
& $-0.0155^{***}$ \\
& $(0.0004)$ \\

Verified
& $-0.0060$ \\
& $(0.0043)$ \\

Intercept
& $0.1450^{***}$ \\
& $(0.0037)$ \\

\hline
Country Fixed Effects & Yes \\
Observations & 84,206 \\
$R^2$ & 0.032 \\
\hline
\multicolumn{2}{l}{Standard errors in parentheses (OLS).} \\
\end{tabular}
\caption{User-level suspension model (country sample) with country fixed effects.}
\label{tab_si:user_suspension_country_reg}

\end{table}

%% file: tables/user_level_suspension_reg_interaction_lang.tex
\begin{table}[H]
\centering
\small
\begin{tabular}{lc}
\hline
 & User Suspended \\
\hline
User Posted Hate (English baseline)
& 0.0170 \\
& (0.0459) \\

\textbf{Language Effects (vs English)} & \\

Arabic & 0.1234$^{***}$ \\
German & -0.0606$^{***}$ \\
Spanish & -0.0692$^{***}$ \\
French & -0.0648$^{***}$ \\
Indonesian & -0.0588$^{***}$ \\
Portuguese & -0.0667$^{***}$ \\
Turkish & 0.0001 \\

\textbf{Hate $\times$ Language} & \\

Arabic & -0.1189$^{*}$ \\
German & 0.0189 \\
Spanish & 0.0019 \\
French & 0.0242 \\
Indonesian & 0.1161 \\
Portuguese & -0.0103 \\
Turkish & -0.0700 \\

\hline
Joint Wald (all $\times$language interactions $=0$) & $F_{7}=3.70$, $p<0.001$ \\
Observations & 213,744 \\
$R^2$ & 0.187 \\
\hline
\multicolumn{2}{l}{\footnotesize $^{***}p<0.001$, $^{**}p<0.01$, $^{*}p<0.05$, $^{\dagger}p<0.10$.} \\
\end{tabular}
\caption{User-level suspension model with language interactions (heteroskedasticity-robust HC1 standard errors in parentheses). English is the reference. Control variables (log tweet count, log followers, log following, verified status) are included as in table~\ref{tab_si:user_suspension_language_reg} but suppressed for compactness.}
\label{tab_si:language_interaction_suspension}
\end{table}

%% file: tables/user_level_suspension_reg_interaction_country.tex
\begin{table}[H]
\centering
\small
\begin{tabular}{lc}
\hline
 & User Suspended \\
\hline
User Posted Hate (United States baseline)
& 0.0513 \\
& (0.0391) \\

\textbf{Country Effects (vs United States)} & \\

India & -0.0136$^{***}$ \\
Nigeria & -0.0061$^{***}$ \\
Kenya & -0.0186$^{***}$ \\

\textbf{Hate $\times$ Country} & \\

India & 0.0364 \\
Nigeria & -0.0691 \\
Kenya & -0.0179 \\

\hline
Joint Wald (all $\times$country interactions $=0$) & $F_{3}=3.75$, $p=0.010$ \\
Observations & 83,965 \\
$R^2$ & 0.010 \\
\hline
\multicolumn{2}{l}{\footnotesize $^{***}p<0.001$, $^{**}p<0.01$, $^{*}p<0.05$, $^{\dagger}p<0.10$.} \\
\end{tabular}
\caption{User-level suspension model with country interactions (heteroskedasticity-robust HC1 standard errors in parentheses). United States is the reference. Control variables (log tweet count, log followers, log following, verified status) are included as in table~\ref{tab_si:user_suspension_country_reg} but suppressed for compactness.}
\label{tab_si:country_interaction_suspension}
\end{table}

%% file: tables/engagement_reg_lang_claude.tex
\begin{table}[H]
\centering
\small
\begin{tabular}{lcc}
\hline
 & \multicolumn{2}{c}{Log Total Engagement (Language Sample)} \\
 & (1) Hate & (2) Violent vs.\ Non-Violent Hate \\
\hline
\textbf{Hate Indicators} & & \\

Hate
& $-0.18^{*}$
&  \\
& $(0.07)$ &  \\

Violent Hate
&  & $0.62$ \\
&  & $(0.38)$ \\

Non-Violent Hate
&  & $-0.21^{**}$ \\
&  & $(0.08)$ \\

\textbf{Controls} & & \\

Possibly Sensitive
& $0.35^{***}$ & $0.35^{***}$ \\
& $(0.02)$ & $(0.02)$ \\

Verified Author
& $1.35^{***}$ & $1.35^{***}$ \\
& $(0.01)$ & $(0.01)$ \\

Is Reply
& $-1.31^{***}$ & $-1.31^{***}$ \\
& $(0.01)$ & $(0.01)$ \\

Scam
& $1.65^{***}$ & $1.65^{***}$ \\
& $(0.06)$ & $(0.06)$ \\

Adult Content
& $0.06^{**}$ & $0.06^{**}$ \\
& $(0.02)$ & $(0.02)$ \\

\hline
Language Fixed Effects & Yes & Yes \\
Observations & 120,000 & 120,000 \\
$R^2$ & 0.41 & 0.41 \\
\hline
\multicolumn{3}{l}{Standard errors in parentheses (OLS).} \\
\multicolumn{3}{l}{$^{***}p<0.001$, $^{**}p<0.01$, $^{*}p<0.05$, $^{\dagger}p<0.10$.}
\end{tabular}
\caption{Linear models predicting log total engagement in the engagement-weighted language sample with language fixed effects. Engagement is defined as $\log(1 + \text{likes} + \text{replies} + \text{retweets} + \text{quotes})$. Column (1) uses an aggregate Twitter-defined hate indicator; Column (2) separates violent and non-violent hate.}
\label{tab_si:engagement_language}
\end{table}

%% file: tables/engagement_reg_country_claude.tex
\begin{table}[H]
\centering
\small
\begin{tabular}{lcc}
\hline
 & \multicolumn{2}{c}{Log Total Engagement (Country Sample)} \\
 & (1) Hate & (2) Violent vs.\ Non-Violent Hate \\
\hline
\textbf{Hate Indicators} & & \\

Hate
& $0.08$
&  \\
& $(0.09)$ &  \\

Violent Hate
&  & $0.32$ \\
&  & $(0.43)$ \\

Non-Violent Hate
&  & $0.07$ \\
&  & $(0.09)$ \\

\textbf{Controls} & & \\

Possibly Sensitive
& $0.31^{***}$ & $0.31^{***}$ \\
& $(0.04)$ & $(0.04)$ \\

Verified Author
& $1.35^{***}$ & $1.35^{***}$ \\
& $(0.01)$ & $(0.01)$ \\

Is Reply
& $-0.71^{***}$ & $-0.71^{***}$ \\
& $(0.01)$ & $(0.01)$ \\

Scam
& $0.66^{***}$ & $0.66^{***}$ \\
& $(0.11)$ & $(0.11)$ \\

Adult Content
& $0.08^{\dagger}$ & $0.08^{\dagger}$ \\
& $(0.05)$ & $(0.05)$ \\

\hline
Country Fixed Effects & Yes & Yes \\
Observations & 60,000 & 60,000 \\
$R^2$ & 0.29 & 0.29 \\
\hline
\multicolumn{3}{l}{Standard errors in parentheses (OLS).} \\
\multicolumn{3}{l}{$^{***}p<0.001$, $^{**}p<0.01$, $^{*}p<0.05$, $^{\dagger}p<0.10$.}
\end{tabular}
\caption{Linear models predicting log total engagement in the engagement-weighted country sample (US, India, Nigeria, Kenya) with country fixed effects. Engagement is defined as $\log(1 + \text{likes} + \text{replies} + \text{retweets} + \text{quotes})$. Column (1) uses an aggregate Twitter-defined hate indicator; Column (2) separates violent and non-violent hate.}
\label{tab_si:engagement_country}
\end{table}

%% file: tables/engagement_removal_interaction_claude.tex
\begin{table}[H]
\centering
\small
\begin{tabular}{lcc}
\hline
 & \multicolumn{2}{c}{Tweet Removed} \\
 & (1) Language Sample & (2) Country Sample \\
\hline
\textbf{Key Variables} & & \\

Hate
& $0.17^{***}$ & $0.08$ \\
& $(0.05)$ & $(0.05)$ \\

Log Total Engagement
& $0.02^{***}$ & $0.02^{***}$ \\
& $(0.00)$ & $(0.00)$ \\

Hate $\times$ Log Total Engagement
& $-0.04^{*}$ & $-0.02$ \\
& $(0.02)$ & $(0.02)$ \\

\textbf{Controls} & & \\

Possibly Sensitive
& $0.03^{***}$ & $0.15^{***}$ \\
& $(0.01)$ & $(0.01)$ \\

Verified Author
& $-0.16^{***}$ & $-0.11^{***}$ \\
& $(0.00)$ & $(0.00)$ \\

Is Reply
& $-0.03^{***}$ & $0.01^{**}$ \\
& $(0.00)$ & $(0.00)$ \\

Scam
& $0.28^{***}$ & $0.28^{***}$ \\
& $(0.03)$ & $(0.04)$ \\

Adult Content
& $0.34^{***}$ & $0.13^{***}$ \\
& $(0.01)$ & $(0.02)$ \\

\hline
Intercept
& $0.25^{***}$ & $0.08^{***}$ \\
& $(0.00)$ & $(0.00)$ \\

\hline
Fixed Effects & Yes (language) & Yes (country) \\
Observations & 120,000 & 60,000 \\
$R^2$ & 0.13 & 0.03 \\
\hline
\multicolumn{3}{l}{Standard errors in parentheses (OLS, HC3).} \\
\multicolumn{3}{l}{$^{***}p<0.001$, $^{**}p<0.01$, $^{*}p<0.05$, $^{\dagger}p<0.10$.}
\end{tabular}
\caption{Linear probability models predicting tweet removal as a function of Twitter-defined hate, early engagement, and their interaction, estimated on engagement-weighted samples. Column (1) uses the language sample (eight languages) with language fixed effects; Column (2) uses the country sample (US, India, Nigeria, Kenya) with country fixed effects. Log Total Engagement is $\log(1 + \text{likes} + \text{replies} + \text{retweets} + \text{quotes})$ measured ten minutes after posting. The interaction term captures whether the association between hate and removal varies with engagement level; figure~\ref{fig:low_enforcement}C plots the implied marginal effect of engagement on hate removal probability across the engagement distribution.}
\label{tab_si:engagement_removal_interaction}
\end{table}

%% file: tables/logistic_ame_lang_claude.tex
\begin{table}[H]
\centering
\small
\begin{tabular}{lcccc}
\hline
 & \multicolumn{4}{c}{Tweet Removed (Language Sample)} \\
 & (1) Hate & (2) Violent vs.\ Non-Violent Hate & (3) + User Controls & (4) + Policy Categories \\
\hline
\textbf{Hate Indicators (vs.\ Neutral)} & & & & \\

$\quad$ Hate
& $0.03^{*}$ &  &  &  \\
& $(0.01)$ &  &  &  \\

$\quad$ Violent Hate
&  & $0.02$ & $0.00$ & $-0.04$ \\
&  & $(0.05)$ & $(0.04)$ & $(0.08)$ \\

$\quad$ Non-Violent Hate
&  & $0.03^{*}$ & $0.03^{*}$ & $0.02$ \\
&  & $(0.01)$ & $(0.01)$ & $(0.02)$ \\

\textbf{Tweet-Level Controls} & & & & \\

$\quad$ Retweet Count (log)
& $0.02^{***}$ & $0.02^{***}$ & $0.02^{***}$ & $0.02^{***}$ \\
& $(0.00)$ & $(0.00)$ & $(0.00)$ & $(0.00)$ \\

$\quad$ Possibly Sensitive
& $0.04^{***}$ & $0.04^{***}$ & $0.07^{***}$ & $0.04^{***}$ \\
& $(0.01)$ & $(0.01)$ & $(0.01)$ & $(0.01)$ \\

$\quad$ Is Reply
& $-0.10^{***}$ & $-0.10^{***}$ & $-0.06^{***}$ & $-0.06^{***}$ \\
& $(0.00)$ & $(0.00)$ & $(0.00)$ & $(0.00)$ \\

\textbf{User-Level Controls} & & & & \\

$\quad$ Verified Author
&  &  & $-0.19^{***}$ & $-0.17^{***}$ \\
&  &  & $(0.02)$ & $(0.03)$ \\

$\quad$ Followers (log)
&  &  & $-0.01^{***}$ & $-0.01^{***}$ \\
&  &  & $(0.00)$ & $(0.00)$ \\

$\quad$ Following (log)
&  &  & $-0.01^{***}$ & $-0.01^{***}$ \\
&  &  & $(0.00)$ & $(0.00)$ \\

$\quad$ User Tweet Count (log)
&  &  & $0.00^{***}$ & $0.00^{***}$ \\
&  &  & $(0.00)$ & $(0.00)$ \\

$\quad$ Account Age (log)
&  &  & $-0.03^{***}$ & $-0.03^{***}$ \\
&  &  & $(0.00)$ & $(0.00)$ \\

\textbf{Additional Policy Categories} & & & & \\

$\quad$ Scam
&  &  &  & $0.18^{***}$ \\
&  &  &  & $(0.02)$ \\

$\quad$ Adult Content
&  &  &  & $0.10^{***}$ \\
&  &  &  & $(0.01)$ \\

\hline
Language Fixed Effects & Yes & Yes & Yes & Yes \\
Observations
& 240,000 & 240,000 & 240,000 & 80,000 \\

\hline
\multicolumn{5}{l}{$^{***}p<0.001$, $^{**}p<0.01$, $^{*}p<0.05$, $^{\dagger}p<0.10$.} \\
\end{tabular}
\caption{Robustness check: logistic regression predicting tweet removal (language sample) with language fixed effects, reporting average marginal effects with standard errors (delta method) in parentheses. Specifications match Columns (1)--(4) of table~\ref{tab_si:tweet_removal_language_reg}.}
\label{tab_si:robustness_1}
\end{table}

%% file: tables/logistic_ame_country_claude.tex
\begin{table}[H]
\centering
\small
\begin{tabular}{lccc}
\hline
 & \multicolumn{3}{c}{Tweet Removed (Country Sample)} \\
 & (1) Hate & (2) Violent vs.\ Non-Violent Hate & (3) + User Controls \\
\hline
\textbf{Hate Indicators (vs.\ Neutral)} & & & \\

$\quad$ Hate
& $0.05^{***}$ &  &  \\
& $(0.01)$ &  &  \\

$\quad$ Violent Hate
&  & $-0.08$ & $-0.07$ \\
&  & $(0.09)$ & $(0.09)$ \\

$\quad$ Non-Violent Hate
&  & $0.05^{***}$ & $0.05^{***}$ \\
&  & $(0.01)$ & $(0.01)$ \\

\textbf{Tweet-Level Controls} & & & \\

$\quad$ Retweet Count (log)
& $-0.00$ & $-0.00$ & $0.01^{***}$ \\
& $(0.00)$ & $(0.00)$ & $(0.00)$ \\

$\quad$ Possibly Sensitive
& $0.05^{***}$ & $0.05^{***}$ & $0.05^{***}$ \\
& $(0.01)$ & $(0.01)$ & $(0.01)$ \\

$\quad$ Is Reply
& $-0.01^{***}$ & $-0.01^{***}$ & $-0.02^{***}$ \\
& $(0.00)$ & $(0.00)$ & $(0.00)$ \\

\textbf{User-Level Controls} & & & \\

$\quad$ Verified Author
&  &  & $-0.12^{***}$ \\
&  &  & $(0.01)$ \\

$\quad$ Followers (log)
&  &  & $-0.00$ \\
&  &  & $(0.00)$ \\

$\quad$ Following (log)
&  &  & $-0.01^{***}$ \\
&  &  & $(0.00)$ \\

$\quad$ User Tweet Count (log)
&  &  & $-0.01^{***}$ \\
&  &  & $(0.00)$ \\

$\quad$ Account Age (log)
&  &  & $-0.01^{***}$ \\
&  &  & $(0.00)$ \\

\hline
Country Fixed Effects & Yes & Yes & Yes \\
Observations
& 120,000 & 120,000 & 120,000 \\

\hline
\multicolumn{4}{l}{$^{***}p<0.001$, $^{**}p<0.01$, $^{*}p<0.05$, $^{\dagger}p<0.10$.} \\
\end{tabular}
\caption{Robustness check: logistic regression predicting tweet removal (country sample) with country fixed effects, reporting average marginal effects with standard errors (delta method) in parentheses. Specifications match Columns (1)--(3) of table~\ref{tab_si:tweet_removal_country_reg}. Column (4) (policy categories) is omitted because the policy-annotated country subsample contains only 8 violent hate tweets, so the logit fails to converge.}
\label{tab_si:robustness_2}
\end{table}

%% file: tables/clustered_se_lang_claude.tex
\begin{table}[H]
\centering
\small
\begin{tabular}{lcccc}
\hline
 & \multicolumn{4}{c}{Tweet Removed (Language Sample)} \\
 & (1) Hate & (2) Violent vs.\ Non-Violent Hate & (3) + User Controls & (4) + Policy Categories \\
\hline
\textbf{Hate Indicators} & & & & \\

Hate
& $0.03^{\dagger}$ &  &  &  \\
& $(0.02)$ &  &  &  \\

Violent Hate
&  & $0.01$ & $0.00$ & $-0.06$ \\
&  & $(0.05)$ & $(0.06)$ & $(0.08)$ \\

Non-Violent Hate
&  & $0.03^{\dagger}$ & $0.02$ & $0.01$ \\
&  & $(0.02)$ & $(0.02)$ & $(0.03)$ \\

\textbf{Tweet-Level Controls} & & & & \\

Retweet Count (log)
& $0.04^{***}$ & $0.04^{***}$ & $0.05^{***}$ & $0.04^{***}$ \\
& $(0.00)$ & $(0.00)$ & $(0.00)$ & $(0.01)$ \\

Possibly Sensitive
& $0.05^{***}$ & $0.05^{***}$ & $0.08^{***}$ & $0.04^{*}$ \\
& $(0.01)$ & $(0.01)$ & $(0.01)$ & $(0.02)$ \\

Is Reply
& $-0.10^{***}$ & $-0.10^{***}$ & $-0.07^{***}$ & $-0.07^{***}$ \\
& $(0.00)$ & $(0.00)$ & $(0.00)$ & $(0.00)$ \\

\textbf{User-Level Controls} & & & & \\

Verified Author
&  &  & $-0.02$ & $-0.01$ \\
&  &  & $(0.01)$ & $(0.01)$ \\

Followers (log)
&  &  & $-0.01^{***}$ & $-0.01^{***}$ \\
&  &  & $(0.00)$ & $(0.00)$ \\

Following (log)
&  &  & $-0.02^{***}$ & $-0.02^{***}$ \\
&  &  & $(0.00)$ & $(0.00)$ \\

User Tweet Count (log)
&  &  & $0.01^{***}$ & $0.01^{***}$ \\
&  &  & $(0.00)$ & $(0.00)$ \\

Account Age (log)
&  &  & $-0.04^{***}$ & $-0.04^{***}$ \\
&  &  & $(0.00)$ & $(0.00)$ \\

\textbf{Additional Policy Categories} & & & & \\

Scam
&  &  &  & $0.29^{***}$ \\
&  &  &  & $(0.05)$ \\

Adult Content
&  &  &  & $0.17^{***}$ \\
&  &  &  & $(0.02)$ \\

\hline
Intercept & $0.45^{***}$ & $0.45^{***}$ & $0.74^{***}$ & $0.73^{***}$ \\
& $(0.00)$ & $(0.00)$ & $(0.00)$ & $(0.01)$ \\
\hline
Language Fixed Effects & Yes & Yes & Yes & Yes \\
Observations & 240,000 & 240,000 & 240,000 & 80,000 \\
$R^2$ & 0.08 & 0.08 & 0.17 & 0.18 \\
\hline
\multicolumn{5}{l}{Standard errors clustered at the user level in parentheses (OLS).} \\
\multicolumn{5}{l}{$^{***}p<0.001$, $^{**}p<0.01$, $^{*}p<0.05$, $^{\dagger}p<0.10$.}
\end{tabular}
\caption{Replication of the main tweet-level removal models for the language sample (table~\ref{tab_si:tweet_removal_language_reg}) with standard errors clustered at the user level. Specifications match Columns (1)--(4) of the main LPM table; only the variance estimator differs.}
\label{tab_si:robustness_clustered_lang}
\end{table}

%% file: tables/clustered_se_country_claude.tex
\begin{table}[H]
\centering
\small
\begin{tabular}{lccc}
\hline
 & \multicolumn{3}{c}{Tweet Removed (Country Sample)} \\
 & (1) Hate & (2) Violent vs.\ Non-Violent Hate & (3) + User Controls \\
\hline
\textbf{Hate Indicators} & & & \\

Hate
& $0.06^{**}$ &  &  \\
& $(0.02)$ &  &  \\

Violent Hate
&  & $-0.06$ & $-0.06$ \\
&  & $(0.04)$ & $(0.05)$ \\

Non-Violent Hate
&  & $0.07^{**}$ & $0.06^{**}$ \\
&  & $(0.02)$ & $(0.02)$ \\

\textbf{Tweet-Level Controls} & & & \\

Retweet Count (log)
& $-0.00$ & $-0.00$ & $0.01$ \\
& $(0.01)$ & $(0.01)$ & $(0.01)$ \\

Possibly Sensitive
& $0.08^{***}$ & $0.08^{***}$ & $0.07^{***}$ \\
& $(0.01)$ & $(0.01)$ & $(0.01)$ \\

Is Reply
& $-0.02^{***}$ & $-0.02^{***}$ & $-0.02^{***}$ \\
& $(0.00)$ & $(0.00)$ & $(0.00)$ \\

\textbf{User-Level Controls} & & & \\

Verified Author
&  &  & $-0.04^{***}$ \\
&  &  & $(0.01)$ \\

Followers (log)
&  &  & $-0.00$ \\
&  &  & $(0.00)$ \\

Following (log)
&  &  & $-0.01^{**}$ \\
&  &  & $(0.00)$ \\

User Tweet Count (log)
&  &  & $-0.01^{***}$ \\
&  &  & $(0.00)$ \\

Account Age (log)
&  &  & $-0.02^{***}$ \\
&  &  & $(0.00)$ \\

\hline
Intercept & $0.12^{***}$ & $0.12^{***}$ & $0.35^{***}$ \\
& $(0.00)$ & $(0.00)$ & $(0.01)$ \\
\hline
Country Fixed Effects & Yes & Yes & Yes \\
Observations & 120,000 & 120,000 & 120,000 \\
$R^2$ & 0.00 & 0.00 & 0.03 \\
\hline
\multicolumn{4}{l}{Standard errors clustered at the user level in parentheses (OLS).} \\
\multicolumn{4}{l}{$^{***}p<0.001$, $^{**}p<0.01$, $^{*}p<0.05$, $^{\dagger}p<0.10$.}
\end{tabular}
\caption{Replication of the main tweet-level removal models for the country sample (table~\ref{tab_si:tweet_removal_country_reg}) with standard errors clustered at the user level. Specifications match Columns (1)--(3) of the main LPM table; only the variance estimator differs. Column (4) (policy categories) is omitted because the policy-annotated country subsample contains only 8 violent hate tweets, yielding numerically unstable cluster-robust standard errors on the violent hate coefficient.}
\label{tab_si:robustness_clustered_country}
\end{table}

%% file: tables/scam_adult_agreement_sensitivity.tex
\begin{table}[H]
\centering
\small
\begin{tabular}{lccc}
\hline
 & \multicolumn{3}{c}{Annotator-agreement threshold} \\
 & $\geq$1/3 (any) & $\geq$2/3 (majority) & 3/3 (unanimous) \\
\hline
\textbf{Varying the scam threshold (adult held at majority)} & & & \\

$\quad$ Scam coefficient
& 0.10$^{***}$ & 0.29$^{***}$ & 0.16$^{**}$ \\
& (0.01) & (0.03) & (0.06) \\

$\quad$ N tweets flagged as scam
& 955 & 148 & 34 \\

$\quad$ Violent hate (for reference)
& -0.06 & -0.06 & -0.06 \\
$\quad$ Non-violent hate (for reference)
& 0.01 & 0.01 & 0.01 \\
$\quad$ Adult content (majority, for reference)
& 0.17$^{***}$ & 0.17$^{***}$ & 0.17$^{***}$ \\

\hline
\textbf{Varying the adult threshold (scam held at majority)} & & & \\

$\quad$ Adult coefficient
& 0.13$^{***}$ & 0.17$^{***}$ & 0.21$^{***}$ \\
& (0.01) & (0.01) & (0.01) \\

$\quad$ N tweets flagged as adult
& 2,037 & 1,268 & 797 \\

\hline
\multicolumn{4}{l}{$^{***}p<0.001$, $^{**}p<0.01$, $^{*}p<0.05$, $^{\dagger}p<0.10$.} \\
\end{tabular}
\caption{Sensitivity of the policy-category removal coefficients to the annotator-agreement threshold used to construct the scam and adult-content indicators (language sample, $N=80{,}000$ joint annotated subset). The headline specification (column~4 of table~\ref{tab_si:tweet_removal_language_reg}) uses majority agreement ($\geq$2/3 annotators). The table additionally reports the any-annotator ($\geq$1/3) and unanimous (3/3) thresholds for each indicator, with the violent and non-violent hate coefficients shown for reference. Standard errors in parentheses (OLS). All specifications include the controls and language fixed effects from column~(4) of table~\ref{tab_si:tweet_removal_language_reg}, and only the scam or adult threshold varies across columns.}
\label{tab_si:scam_adult_agreement_sensitivity}
\end{table}

%% file: tables/scam_adult_channel_decomp.tex
\begin{table}[H]
\centering
\footnotesize
\begin{tabular}{lcccc}
\multicolumn{5}{l}{\textbf{Panel A. Removal channel (dependent variable swapped; column-4 specification)}} \\
\hline
 & Removed & Suspended & Deleted & Deactivated \\
 & (all) & (account-level) & (tweet-level) & (user-initiated) \\
\hline
Violent hate & -6.1 & -10.7$^{*}$ & +4.6 & -0.8 \\
 & (7.7) & (5.5) & (6.0) & (1.7) \\
Non-violent hate & +0.9 & -1.1 & +2.0 & -0.3 \\
 & (2.4) & (1.7) & (1.9) & (0.6) \\
Scam & +29.0$^{***}$ & +35.1$^{***}$ & -6.0$^{**}$ & +0.1 \\
 & (2.9) & (2.1) & (2.3) & (0.7) \\
Adult content & +17.1$^{***}$ & +16.8$^{***}$ & +0.3 & +0.7$^{**}$ \\
 & (1.0) & (0.7) & (0.8) & (0.2) \\
\hline
\multicolumn{5}{l}{\emph{Among category-positive tweets, count removed via each channel:}} \\
\quad Scam ($n=147$) & & 78 & 5 & 1 \\
\quad Adult ($n=1{,}269$) & & 485 & 150 & 17 \\
\quad Hate ($n=231$) & & 18 & 27 & 1 \\
Observations & 80,000 & 80,000 & 80,000 & 80,000 \\
\hline
\\
\multicolumn{5}{l}{\textbf{Panel B. Robustness to bot-like account controls (dependent variable: removed)}} \\
\hline
 & \multicolumn{3}{c}{Control set} & Attenuation \\
 & (i) Tweet & (ii) +User & (iii) +Bot & (i)$\to$(iii) \\
 & controls & controls & signals & \\
\hline
Scam & +38.0$^{***}$ & +29.0$^{***}$ & +27.5$^{***}$ & +28\% \\
 & (3.1) & (2.9) & (2.9) & \\
Adult content & +23.3$^{***}$ & +17.1$^{***}$ & +16.9$^{***}$ & +27\% \\
 & (1.1) & (1.0) & (1.0) & \\
Violent hate & -6.2 & -6.1 & -5.7 & --- \\
 & (8.1) & (7.7) & (7.7) & \\
Non-violent hate & +0.8 & +0.9 & +1.3 & --- \\
 & (2.6) & (2.4) & (2.4) & \\
Observations & 80,000 & 80,000 & 80,000 & \\
\hline
\\
\multicolumn{5}{l}{\textbf{Panel C. Adult content $\times$ sensitive-media flag (dependent variable: removed)}} \\
\hline
Adult (text-only, not sensitive-flagged) & \multicolumn{4}{l}{+17.6$^{***}$ \quad (1.1)} \\
Adult $\times$ sensitive-media & \multicolumn{4}{l}{-6.6$^{\dagger}$ \quad (3.7)} \\
Adult (sensitive-flagged; sum) & \multicolumn{4}{l}{+11.1$^{**}$ \quad (3.6)} \\
\quad adult tweets sensitive-flagged / text-only & \multicolumn{4}{l}{119 (rm 41\%) / 1150 (rm 51\%)} \\
\hline
\multicolumn{5}{l}{\footnotesize $^{***}p<0.001$, $^{**}p<0.01$, $^{*}p<0.05$, $^{\dagger}p<0.10$.} \\
\end{tabular}
\caption{Decomposition of the scam and adult-content removal advantage over hate by enforcement channel and account characteristics (language sample, $N=80{,}000$ annotated subset), holding the column~(4) specification of table~\ref{tab_si:tweet_removal_language_reg} fixed. Models are linear probability models with language fixed effects, and coefficients are in percentage points with standard errors in parentheses (OLS). Panel~A splits removal into its mutually exclusive channels, where ``Removed'' $=$ ``Suspended'' $+$ ``Deleted''. Panel~B adds bot-like account signals (log follower-to-following ratio and log lifetime tweets per day). Panel~C compares adult-content removal among tweets with and without the platform's sensitive-media flag.}
\label{tab_si:scam_adult_channel_decomp}
\end{table}

%% file: tables/si_attribution_descriptive.tex
\begin{table}[H]
\centering
\footnotesize
\begin{tabular}{lrrrrrrrr}
\hline
 & $n$ & Median & Mean share & Mean share & Mean share & Median & Median & Median \\
 & users & other & hate & adult & scam & age & follow- & follow- \\
 &  & tweets & (other) & (other) & (other) & (days) & ers & ing \\
\hline
\multicolumn{9}{l}{\textit{Group A (scam/adult-annotated)}} \\
\quad Suspended: No & 2,536 & 14 & 0.254 & 0.093 & 0.007 & 417 & 630 & 174 \\
\quad Suspended: Yes & 1,234 & 50 & 0.348 & 0.126 & 0.011 & 279 & 22 & 27 \\
\multicolumn{9}{l}{\textit{Group B (hate-annotated)}} \\
\quad Suspended: No & 817 & 18 & 0.268 & 0.004 & 0.000 & 874 & 275 & 310 \\
\quad Suspended: Yes & 85 & 48 & 0.320 & 0.006 & 0.000 & 221 & 143 & 214 \\
\multicolumn{9}{l}{\textit{Group C (neutral controls)}} \\
\quad Suspended: No & 4,221 & 15 & 0.109 & 0.005 & 0.002 & 1035 & 401 & 364 \\
\quad Suspended: Yes & 381 & 116 & 0.130 & 0.029 & 0.232 & 112 & 21 & 20 \\
\hline
\end{tabular}
\caption{User-level descriptive statistics by annotation group and suspension status, restricted to users in language buckets (en, ar, es, pt, in, tr, fr, de). Group A users had their single annotated tweet labeled as scam or adult, Group B users had it labeled as Twitter-defined hate, and Group C users had it labeled as neutral and were stratified-matched to the union of A and B on language. Shares are computed over each user's non-annotated tweets posted in the 24-hour window. The hate share uses the per-language Perspective \textsc{Identity\_Attack} threshold (German: \texttt{jagoldz-gahd}), and adult and scam shares use lexical and regex classifiers tuned to the dataset.}
\label{tab_si:attribution_descriptive}
\end{table}

%% file: tables/si_attribution_regression.tex
\begin{table}[H]
\centering
\footnotesize
\begin{tabular}{lc c c c}
\hline
 & (1) & (2) & (3) & (4) \\
 & Content & + Group & + Metadata & + Language \\
\hline
Share hate (other tweets) & 1.461$^{***}$ & 0.939$^{***}$ & 0.544$^{***}$ & 0.516$^{***}$ \\
 & (0.098) & (0.104) & (0.124) & (0.128) \\
Share adult (other tweets) & 0.706$^{***}$ & 0.079 & -0.270$^{\dagger}$ & -0.452$^{**}$ \\
 & (0.134) & (0.136) & (0.155) & (0.165) \\
Share scam (other tweets) & 3.812$^{***}$ & 4.336$^{***}$ & 0.837$^{*}$ & 1.122$^{**}$ \\
 & (0.275) & (0.268) & (0.417) & (0.430) \\
Group A (scam/adult-annotated) & --- & 1.709$^{***}$ & 1.439$^{***}$ & 1.517$^{***}$ \\
 &  & (0.071) & (0.082) & (0.084) \\
Group B (hate-annotated) & --- & 0.212 & 0.366$^{**}$ & 0.507$^{***}$ \\
 &  & (0.130) & (0.137) & (0.140) \\
Log other-tweet count & --- & --- & 0.549$^{***}$ & 0.462$^{***}$ \\
 &  &  & (0.024) & (0.025) \\
Log followers & --- & --- & -0.076$^{***}$ & -0.049$^{***}$ \\
 &  &  & (0.013) & (0.014) \\
Log following & --- & --- & -0.236$^{***}$ & -0.212$^{***}$ \\
 &  &  & (0.019) & (0.019) \\
Log tweets/day & --- & --- & -0.272$^{***}$ & -0.199$^{***}$ \\
 &  &  & (0.027) & (0.028) \\
Verified & --- & --- & -2.504$^{*}$ & -2.585$^{*}$ \\
 &  &  & (1.023) & (1.018) \\
\hline
Language FE & No & No & No & Yes \\
Observations & 9,274 & 9,274 & 9,070 & 9,070 \\
Pseudo $R^2$ & 0.069 & 0.151 & 0.292 & 0.315 \\
\hline
\multicolumn{5}{l}{\footnotesize $^{\dagger}p<0.10$, $^{*}p<0.05$, $^{**}p<0.01$, $^{***}p<0.001$.} \\
\end{tabular}
\caption{Logistic regression of user suspension on content posted in non-annotated tweets, annotation-group dummies, and user metadata. Sample restricted to users in language buckets (en, ar, es, pt, in, tr, fr, de). Group~A users had their annotated tweet labeled as scam or adult, Group~B as Twitter-defined hate, and Group~C (the reference) as neutral. Logit coefficients are reported with standard errors in parentheses.}
\label{tab_si:attribution_regression}
\end{table}

%% file: tables/best_models_ap_claude.tex
\begin{table}[H]
\centering
\small
\begin{tabular}{llll}
\hline
\textbf{Context} & \textbf{Language / Country} & \textbf{Best Model} & \textbf{AP} \\
\hline
\multirow{8}{*}{Language}
 & Arabic     & Perspective API & 0.08 \\
 & German     & Perspective API & 0.12 \\
 & English    & Perspective API & 0.11 \\
 & Spanish    & Perspective API & 0.28 \\
 & French     & Perspective API & 0.31 \\
 & Indonesian & Perspective API & 0.08 \\
 & Portuguese & Perspective API & 0.21 \\
 & Turkish    & Perspective API & 0.02 \\
\hline
\multirow{4}{*}{Country}
 & United States & \texttt{Hate-speech-CNERG/dehatebert-mono-english} & 0.07 \\
 & India         & Perspective API & 0.28 \\
 & Nigeria       & Perspective API & 0.05 \\
 & Kenya         & Perspective API & 0.10 \\
\hline
\end{tabular}
\caption{Best-performing model per language and country context, selected by highest average precision (AP) on the unweighted representative HateDay dataset, using \textit{Hate} as the ground-truth label. AP reflects ranking quality across thresholds and is robust to class imbalance. Perspective API scores use the Identity Attack attribute. All models were evaluated with pre-trained weights, without fine-tuning.}
\label{tab_si:best_models_ap}
\end{table}

%% file: figures/si_coverage_results.tex
\begin{figure}[H]
  \centering
  \includegraphics[width=\textwidth]{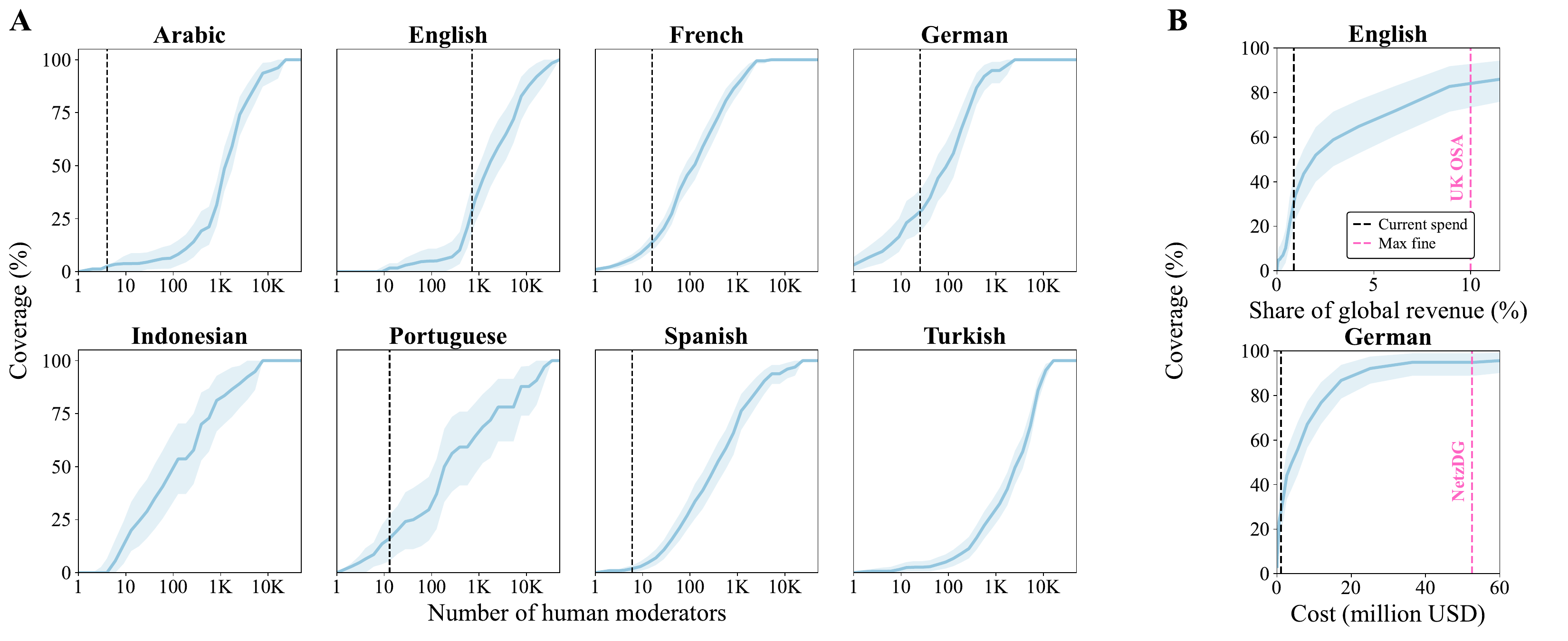}
  \caption{Coverage analogue of Fig.~\ref{fig:human-in-the-loop}. (\textbf{A}) Moderation coverage, the share of hateful tweets surfaced for human review, as a function of the number of hate-speech moderators (log scale), shown separately for each of the eight language communities. Solid lines show median bootstrap estimates. Shaded areas indicate 95\% bootstrap confidence intervals. Black dashed vertical lines mark reported staffing levels in Twitter's August--October 2023 DSA Transparency Report, scaled by the estimated share of moderation activity devoted to hate speech (31\%). Indonesian and Turkish are omitted from this overlay as they are not reported in the DSA filing. (\textbf{B}) Coverage as a function of moderation cost, for English (top) and German (bottom), mirroring Fig.~\ref{fig:human-in-the-loop}B. Cost is expressed as a share of Twitter's 2021 annual revenue (USD 5.08 billion) for English, against the UK Online Safety Act's maximum fine of 10\% of global turnover (pink dashed line), and in absolute USD for German, against NetzDG's maximum fine of EUR 50~million (pink dashed line). Black dashed lines mark the current spend on hate moderation implied by Twitter's reported moderator workforce. Parameters: 3 reviewers per tweet, 100 tweets reviewed per hour. Coverage is estimated on the random (representative) sample.}
  \label{fig:si_coverage_results}
\end{figure}

%% file: tables/engagement_concentration.tex
\begin{table}[H]
\centering
\small
\begin{tabular}{lrrrr}
\hline
\textbf{Language} & \textbf{$N$ hate} & \textbf{No early eng.\ (\%)} & \textbf{Top \% for 80\% eng.} & \textbf{Gini} \\
\hline
Arabic     &  76 & 72.4 & 18.4 & 0.82 \\
English    &  58 & 70.7 & 19.0 & 0.82 \\
French     & 181 & 65.7 &  3.9 & 0.93 \\
German     &  74 & 63.5 & 18.9 & 0.80 \\
Indonesian &  37 & 67.6 & 21.6 & 0.79 \\
Portuguese &  32 & 71.9 & 21.9 & 0.80 \\
Spanish    &  97 & 67.0 & 13.4 & 0.87 \\
Turkish    & 151 & 64.2 & 19.2 & 0.81 \\
\hline
\textbf{Pooled} & 706 & 66.9 & 11.2 & 0.89 \\
\hline
\end{tabular}
\caption{Concentration of hateful engagement by language, computed on the random sample. Early engagement is the total number of interactions within ten minutes of posting. ``Top \% for 80\% eng.'' is the smallest share of hateful tweets, ranked by engagement, that jointly accounts for $80\%$ of hateful engagement. Gini is computed over all hateful tweets in the language, including those with zero engagement.}
\label{tab_si:engagement_concentration}
\end{table}

%% file: tables/sensitivity_params_revised.tex
\begin{table}[H]
\centering
\small
\begin{tabular}{ccrrrr}
\hline
\textbf{Reviewers/} & \textbf{Tweets/} & \textbf{Moderators} & \textbf{English} & \textbf{German} & \textbf{All 8 languages} \\
\textbf{tweet} & \textbf{hour} & \textbf{(all 8)} & \textbf{(\% revenue)} & \textbf{(million USD)} & \textbf{(\% revenue)} \\
\hline
\multicolumn{6}{l}{\textit{Panel A: Avoided hate engagement $\geq$ 80\%}} \\
\hline
1 & 100 & 7,222 & 0.4 & 6.9 & 2.6 \\
1 & 150 & 4,814 & 0.3 & 4.6 & 1.8 \\
1 & 200 & 3,611 & 0.2 & 3.4 & 1.3 \\
\textbf{3} & \textbf{100} & \textbf{21,665} & \textbf{1.3} & \textbf{20.7} & \textbf{7.9} \\
3 & 150 & 14,443 & 0.9 & 13.8 & 5.3 \\
3 & 200 & 10,832 & 0.7 & 10.3 & 4.0 \\
\hline
\multicolumn{6}{l}{\textit{Panel B: Coverage $\geq$ 80\%}} \\
\hline
1 & 100 & 8,885 & 2.7 & 4.5 & 5.5 \\
1 & 150 & 5,923 & 1.8 & 3.0 & 3.7 \\
1 & 200 & 4,442 & 1.4 & 2.3 & 2.8 \\
\textbf{3} & \textbf{100} & \textbf{26,654} & \textbf{8.2} & \textbf{13.6} & \textbf{16.6} \\
3 & 150 & 17,769 & 5.5 & 9.1 & 11.1 \\
3 & 200 & 13,327 & 4.1 & 6.8 & 8.3 \\
\hline
\end{tabular}
\caption{Sensitivity of required moderator counts and annual costs to key operational parameters. Each row shows the total number of moderators across the eight language communities, and the annual cost of reaching 80\% avoided hate engagement (Panel A) or 80\% hate speech coverage (Panel B), for English alone, German alone, and all eight languages combined. English and all-language costs are expressed as a share of Twitter's last audited annual revenue (USD 5.08 billion for fiscal year 2021), and German costs in absolute USD, matching the penalty each regime specifies (Fig.~\ref{fig:human-in-the-loop}B). The number of reviewers per tweet and the review rate are varied around their baseline values (bold). Avoided hate engagement is estimated on the engagement-weighted sample and coverage on the random sample. Moderator requirements and costs scale directly with the number of reviewers per tweet and inversely with the review rate, so the non-baseline rows follow from the baseline simulation.}
\label{tab_si:sensitivity_params}
\end{table}

%% file: tables/wage_assumptions.tex
\begin{table}[H]
\centering
\small
\begin{tabular}{llr}
\hline
\textbf{Language} & \textbf{Reference country} & \textbf{Hourly wage (USD)} \\
\hline
Arabic     & Egypt      &  4.53 \\
English    & United States & 20.00 \\
French     & France     & 15.67 \\
German     & Germany    & 15.15 \\
Indonesian & Indonesia  &  6.96 \\
Portuguese & Brazil     &  9.60 \\
Spanish    & Mexico     & 10.94 \\
Turkish    & Turkey     &  5.88 \\
\hline
\end{tabular}
\caption{Hourly wages used in cost calculations, expressed in USD. Wages are derived by converting a USD 20 baseline to the local purchasing-power equivalent of each language's most populous reference country.}
\label{tab_si:wage_assumptions}
\end{table}